\documentclass[usenatbib,twocolumn]{mn2e}

\usepackage{amsmath}
\usepackage{graphicx}
\usepackage{multicol}
\usepackage{verbatim}

\title{Quantifying substructures in {\it Hubble Frontier Field} clusters: comparison
with $\Lambda CDM$ simulations}

\author[I. Mohammed et al.] {Irshad Mohammed\thanks{irshad@physik.uzh.ch},$^{1,2,3}$
Prasenjit Saha$^{2,3}$, Liliya L. R. Williams$^4$, Jori Liesenborgs$^5$
\newauthor
and Kevin Sebesta$^4$
\\
$^1${Theoretical Astrophysics Group, Fermi National Accelerator Laboratory, Batavia, IL 60510, USA}\\
$^2${Physik-Institut, University of Zurich, Winterthurerstrasse 190, 8057 Zurich, Switzerland}\\
$^3${Institute for Computational Science, University of Zurich, Winterthurerstrasse 190, 8057 Zurich, Switzerland}\\
$^4${School of Physics \& Astronomy, University of Minnesota, 116 Church Street SE, Minneapolis, MN 55455, USA}\\
$^5${Expertisecentrum voor Digitale Media, Universiteit Hasselt, Wetenschapspark 2, B-3590, Diepenbeek, Belgium}
}

\begin{document}
\maketitle


\begin{abstract}

The Hubble Frontier Fields (HFF) are six clusters of galaxies, all showing indications of
recent mergers, which have recently been observed for lensed images.
As such they are the natural laboratories to study the merging history of galaxy clusters.
In this work, we explore the 2D power spectrum of the mass distribution $P_{\rm M}(k)$
as a measure of substructure.
We compare $P_{\rm M}(k)$  of these clusters (obtained using
strong gravitational lensing) to that of $\Lambda$CDM simulated clusters of similar mass.
To compute lensing $P_{\rm M}(k)$,  we produced
free-form lensing mass reconstructions of HFF clusters, without any light traces mass (LTM)
assumption. The inferred power at small scales tends to be larger if (i)~the cluster is
at lower redshift, and/or (ii)~there are deeper observations and hence more lensed images.
In contrast, lens reconstructions assuming LTM show higher power at small
scales even with fewer lensed images; it appears the small scale power in the LTM reconstructions
is dominated by light information, rather than the lensing data.
The average lensing derived $P_{\rm M}(k)$ shows lower power at small scales as compared
to that of simulated clusters at redshift zero, both dark-matter only and hydrodynamical.
The possible reasons are: (i)~the available strong lensing data are limited in their
effective spatial resolution on the mass distribution, (ii)~HFF clusters have yet to build
the small scale power they would have at $z\sim 0$, or (iii)~simulations are
somehow overestimating the small scale power.

\end{abstract}




\section{Introduction}\label{sec:intro}

Clusters of galaxies are the largest self-gravitating objects in the
Universe.  Their ultimate origins must lie in some process of quantum
fluctuations in an expanding universe
\citep{1939Phy.....6..899S,1967RvMP...39..862H,2015PhRvD..91f3505F},
but the earliest observable precursors of galaxy clusters are
fluctuations in the cosmic microwave background.  CMB fluctuations on
the scale of individual clusters would be at $l\sim10^4$, far in the
diffusion-damping tail \citep{1968ApJ...151..459S} and so far barely
accessible observationally \citep{2009ApJ...694.1200R}, but their
subsequent growth through linear gravitational instability is
straightforward.

The early observable proto-clusters
\citep[e.g,][]{2014ApJ...792...15T} are, however, far beyond the baby
fluctuations of the linear regime of gravitational instability --- in
the well-known toy model of spherical collapse, the dynamical outer
boundary of a cluster is at an overdensity of $\simeq 200$ times the
critical density of the Universe.
In this regime some analytical methods based on generalising spherical
collapse are available
\citep{1974ApJ...187..425P,1991ApJ...379..440B,2000MNRAS.318..203S}
but, for the most part, theoretical study depends on numerical
simulations.

With the recent developments in computational resources, it is now possible
to simulate dark matter and gas in cosmological volumes with good
resolution.  In such simulations, it is possible to track the evolution
and mergers of small systems to form large collapsed clusters of
galaxies.  The agreement between simulations and observations have
been improving for various macroscopic properties of galaxies, such as
intergalactic gas, bulge sizes etc.
\citep{2011ApJ...740..102K,2012MNRAS.426.2046A,2012arXiv1206.2838A,
  2014arXiv1407.2600S,2011MNRAS.410.1391A,2013MNRAS.436.3031V,
  2014Natur.509..177V,2014MNRAS.437.1750M,2015MNRAS.446..521S}.  To
study individual objects in more detail, hydrodynamical zoom-in
simulations can be used
\citep{2015MNRAS.446.1939F,2015MNRAS.446.1957F}
in which individual objects are re-simulated at higher resolution
using initial conditions derived from the larger-volume simulation.

A more ambitious comparison between simulations and observations
would be that in terms of the distribution of mass in the clusters.
This is non-trivial for two reasons:

\begin{itemize}
\item First, mass is not an observable; what we observe on the sky is
  light, mass can only be inferred.  Tracing the mass typically
  involves additional assumptions, such as that galaxies sit in the
  potential wells of the dark-matter, light traces mass with some
  scaling parameters, etc.
\item Second, since individual halos, galaxies or clusters of
  galaxies are the outcome of gravitational collapse and various
  baryonic processes of a random field (initial density field), it is
  not possible to compare their mass distribution or clustering
  properties directly. The best one can hope for is to compare them
  statistically.
\end{itemize}

The first problem could be solved with the help of gravitational
lensing, which is sensitive only to the total mass.  Well-known examples of
light not tracing mass, revealed by lensing, are the clusters
ACO~2744 \citep{2011MNRAS.417..333M} and
recently ACO~3827 from \cite{2015MNRAS.449.3393M} \citep[see
  also][]{2011MNRAS.415..448W, 2014MNRAS.439.2651M}.  But simulations of
cluster formation cannot be fitted so as to reproduce detailed
properties of individual clusters.

The second problem could be handled by identifying robust statistical
properties of the clusters.  The simplest quantity is the radial
density profiles.  \cite{2013ApJ...765...24N,2013ApJ...765...25N}
studied the average density profiles of the lensing clusters and
compared them to the simulated clusters. However, this quantity can
only be measured if the cluster is virialised, definitely not for an
ongoing merger.  High-mass clusters in simulations do not usually
virialise until $z\sim0.3$; before that they often show elongation,
multiple cores and many substructures indicating a recent merger, as
do observed lensing clusters.

Another very popular statistic is the mass function; counting the
number of subhalos as a function of their masses
\citep{nat04,nat07,2015ApJ...800...18A}.  However, measuring or even
identifying substructures in lensing clusters has so far been
possible only under the assumption that light traces mass. Without
this assumption, the mass function does not seem a viable statistic
with lensing, especially since lensing gives information about the
sky-projected mass and projection may wash out substructures.

In this work, we propose a different strategy, which is to use a
two-dimensional power spectrum as a basis for comparing lensing
clusters with simulations.  In section \ref{sec:pk}, we define such a
power spectrum, which is normalised differently from the usual
cosmological power spectrum, and has dimensions of mass-squared.  In
section \ref{sec:grale}, we briefly describe GRALE, a
free-form lens-reconstruction technique and code.  In
section~\ref{sec:hff} we apply GRALE to reconstruct the six clusters
of the Hubble Frontier Field from strong-lensing data, and calculate
their power spectra.  Then in section~\ref{sec:comparison} we compare
the clusters with each other and with $\Lambda$CDM simulations, both
dark-matter only and hydrodynamical.  Finally,
section~\ref{sec:discussion} has general discussion and suggestions
for future work.


\section{substructure power spectrum}\label{sec:pk}

Since every cluster is different, comparison of mass maps of observed and simulated clusters is necessarily statistical.  Properties like number of galaxies, 1D density profiles,
concentration, mass function of the structures/substructures, temperature profiles etc.
are useful. However, none of these quantifies the clustering properties of the halo/cluster,
which contains important information about its merging history and evolution.
So it is interesting to study statistically the cluster mass distribution.
The simplest statistic is a two-point function.
Cosmologically, a two-point function gives an excess probability that a local density peak
exists close to another peak, as a function of the separation between the two. It can be well studied
in the Fourier space as the power spectrum. Generally, a power spectrum
$P(k)$ of a 3 dimensional matter density field is defined as:
\begin{equation}
    P(k) = \langle|\tilde{\delta}(k)|^2\rangle
    \label{eqn:pkreal}
\end{equation}
where, $\tilde{\delta}(k)$ is the Fourier transform of the over-density at position corresponding
to the wave-vector $k$.

In this work, we measure the 2D
power spectrum from the projected mass distribution of the halos:
\begin{equation}
    P_{\rm M}(k) = \langle|\tilde{M}_{2D}(k)|^2\rangle
    \label{eqn:pkour}
\end{equation}
where, $\tilde{M}_{2D}(k)$ is the Fourier transform of the 2D projected mass element at position
corresponding to the wave-vector $k$.
\begin{equation}
\tilde M(\vec{k}) = \int \Sigma(\vec{x}) e^{i\vec{k}\cdot\vec{x}} d^2\vec{x}
\end{equation}
This form also gives a natural normalisation of the function
to be the square of the total mass of the halo.

This is not the only way to describe the statistical properties of a density peak, For example,
\cite{2014arXiv1403.2720H}, using the halo model approach, define
\begin{equation}
    P_{\rm 2D}(k) = \int \dfrac{dn}{dM} |\tilde{\kappa}(k)|^2 dM,
\end{equation}
where $P_{\rm 2D}(k)$ is the so-called one-halo term in two dimensions,
$\dfrac{dn}{dM}$ is the differential mass function (number density of halos per unit mass) and
$\tilde{\kappa}(k)$ is the Fourier transform of the convergence (the normalised surface mass density).
The framework is based on the assumptions of virialised halos, and a functional
form for its mass function and radial density profile of each halo. The total contribution to the
power spectrum comes by adding the correlation between different halos (the two-halo term) and
the correlation between matter within the same halo (the one-halo term). Within a halo, there are also
many substructures. If one wants to study the clustering properties of matter within a gravitationally
bound system like a galaxy cluster (or a merging system of galaxies), the two-halo term can be
neglected as all the structures are moving within the gravitational potential of the host and are
indifferent to each other's gravity. However, the one-halo term still exists, and at large scales it
contributes to the Poisson distribution of the substructures which drops at the size of the largest
substructure.
However, this can only apply if the cluster (or the halo) is virialised so that a smooth density
profile can be subtracted in order to see the correlation between the residual field. Now, this is
not a good assumption overall, especially during merging. As most of the massive galaxy
cluster lenses at redshift $>0.3-0.4$ are not virialised systems, this form of the power spectrum
is not intuitive. Therefore, for a general system, including ongoing mergers, recent mergers
and virialised halos, the correlation between the distribution of matter inside the halos
must be studied without such assumptions.

\begin{figure*}
	\centering
	\includegraphics[width=0.32\textwidth]{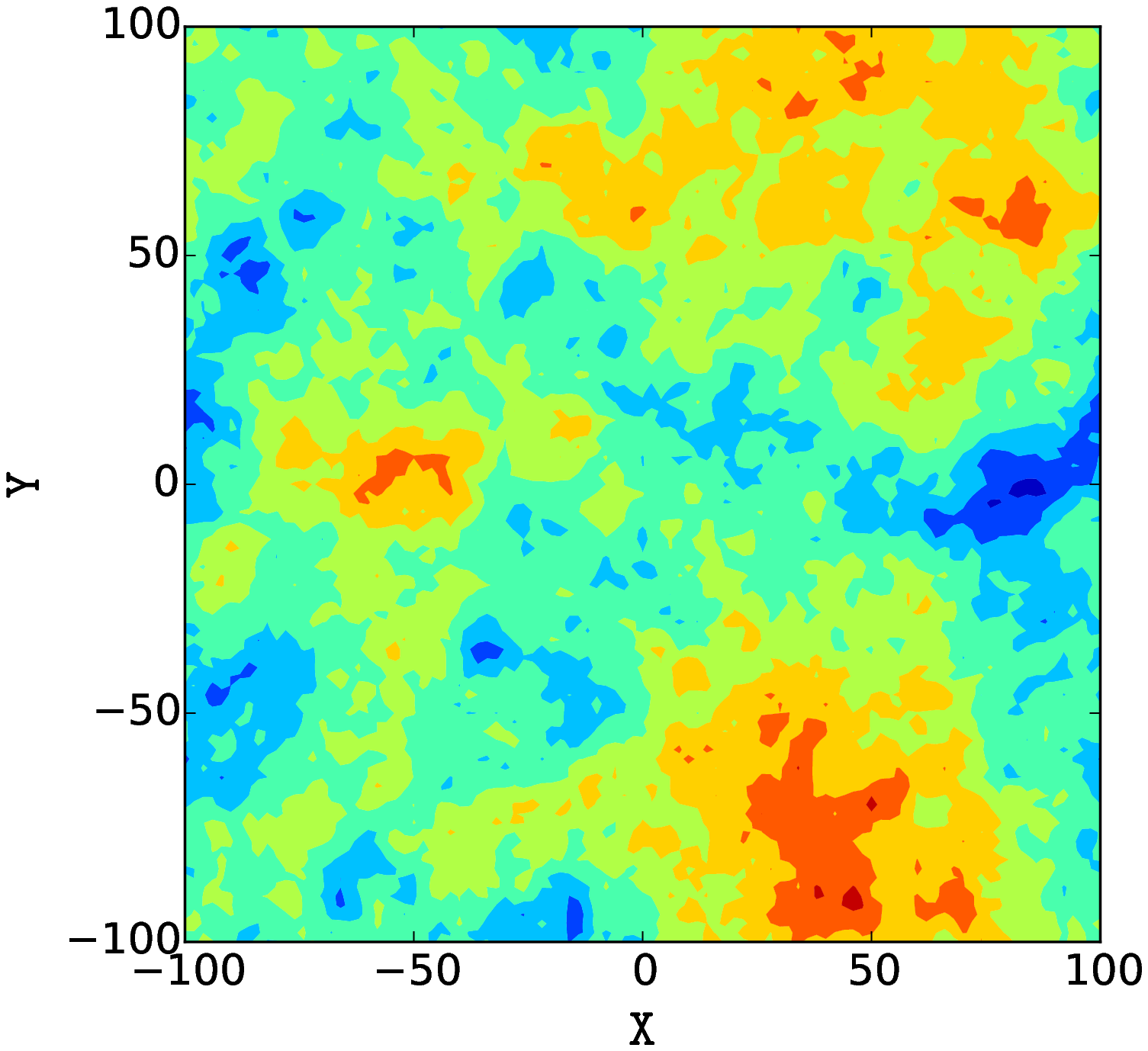}
	\includegraphics[width=0.32\textwidth]{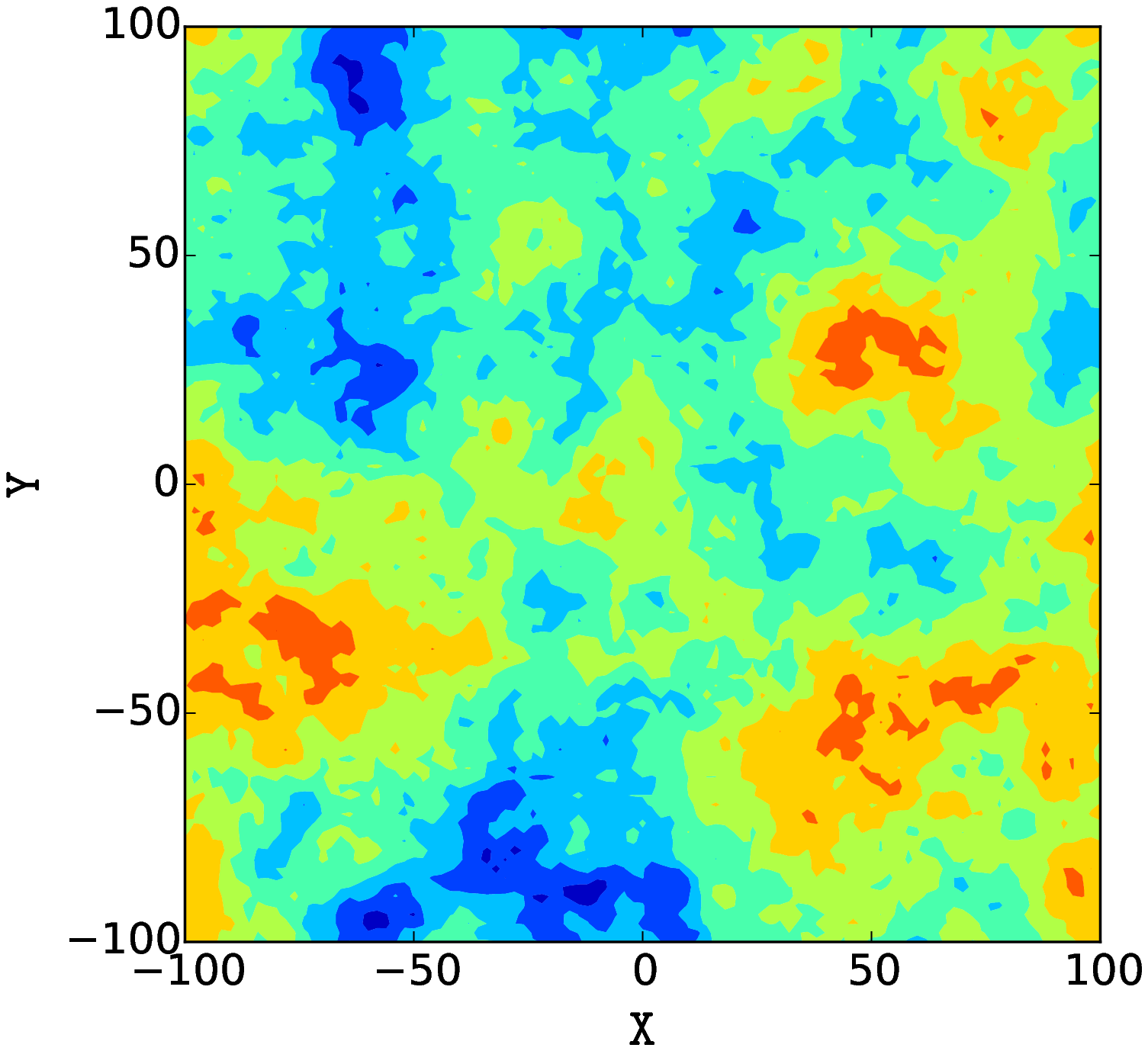}
	\includegraphics[width=0.32\textwidth]{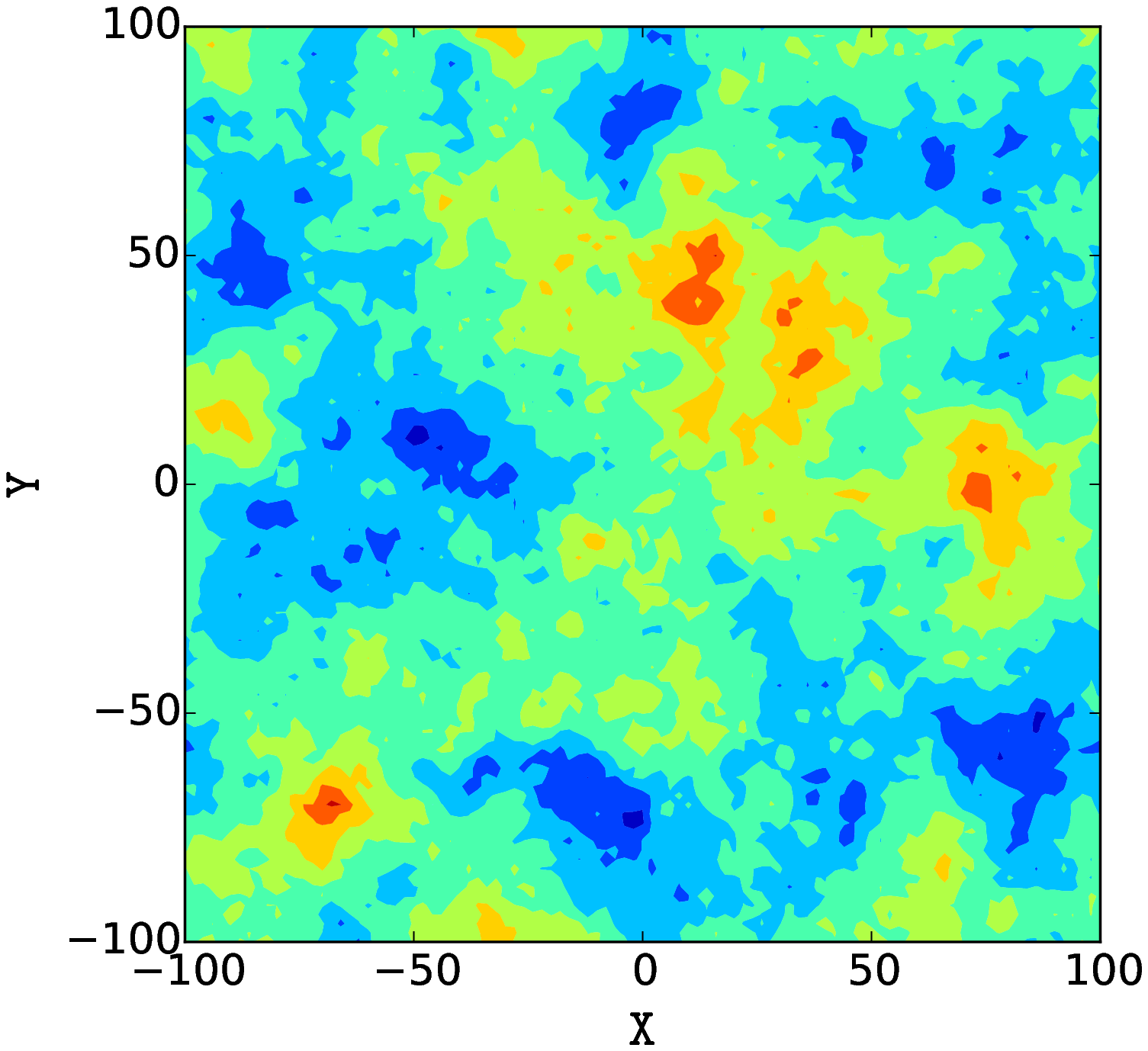}\\
	\includegraphics[width=0.48\textwidth]{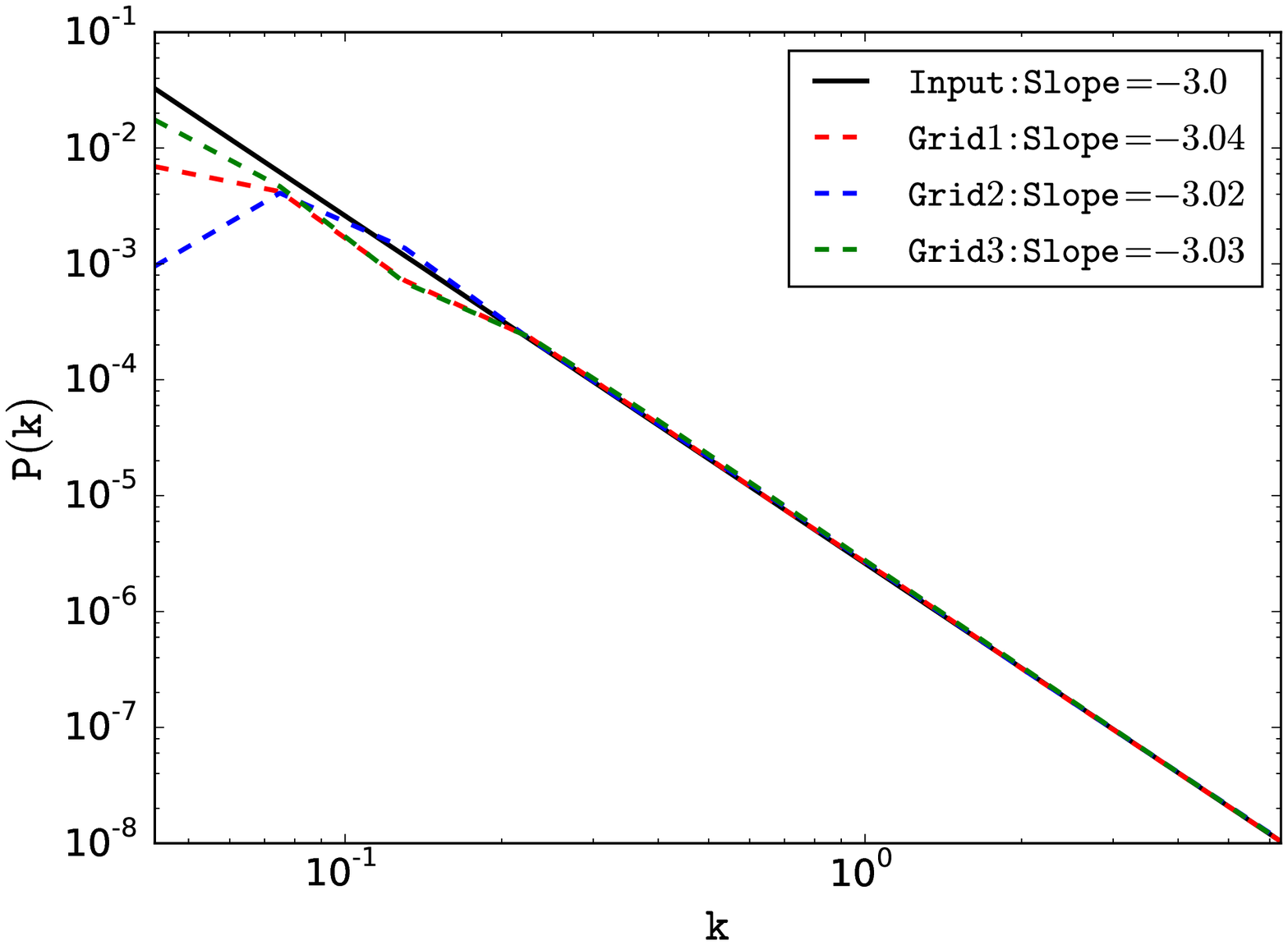}
	\includegraphics[width=0.48\textwidth]{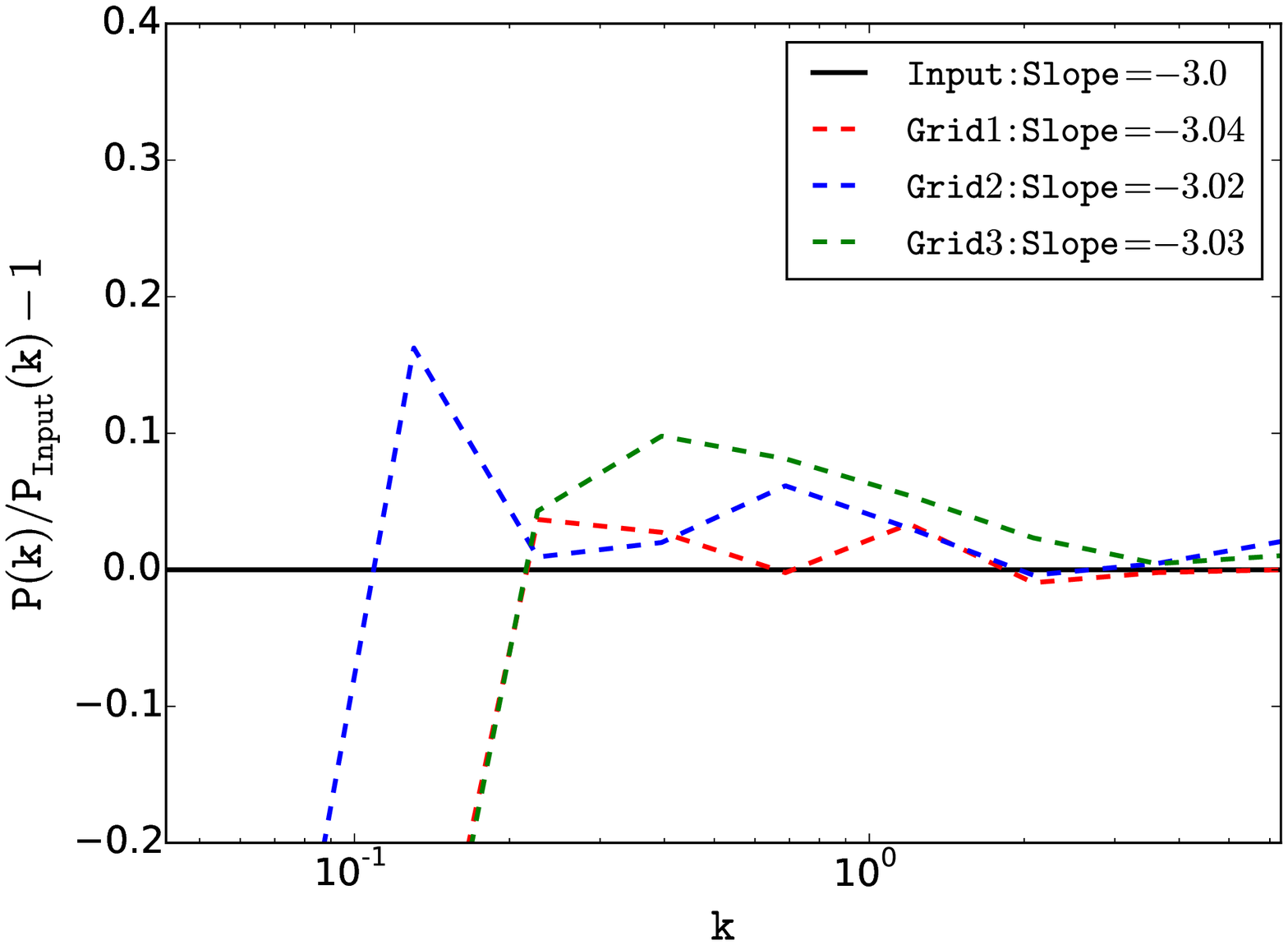}
	\caption{The first three plots (in the top row: Grid1, Grid2, Grid3) are
	the three density map realizations generated
	from the same power spectrum with $k^{-3}$ towards large $k$. The bottom-left panel shows the actual power spectra of each of the grids along with
	the true power spectrum with slope = -3.
	The bottom-right panel shows the
	comparison of the recovered power spectrum from each density grid to the input power.  These maps were only produced to test the numerical computation of the power spectrum, and have no central cluster. While it is possible to generate random fields constrained to have certain features \citep[cf.][]{1991ApJ...380L...5H} these were not needed for this simple test.}
	\label{fig:simgrid}
\end{figure*}

With our definition, if we have a field enclosing only one halo
with its substructures, and we compute $P_{\rm M}(k)$ at scales larger than the size of the halo, for
all those scales one gets the same mass and hence, the power above all those scales will be a constant,
which mimics the Poisson noise of the substructures in the halo model picture. Further this function
drops at scales smaller than the size of the halo,
which is the largest structure in the field, consistent
with the halo model intuition.
For a virialised halo, this function will be smooth and a power law for all scales smaller than
the size of the halo. However, if the halo is not spherically symmetric and in a merging stage
the fluctuations in the function resemble variation in its power at various scales. If one
identifies a large ensemble of halos (let's say in simulations) of nearly the same mass and
the same epoch and averages all the power spectra, one will get an unbiased trend of the clustering
properties of such halos statistically. However, studying individual systems is like studying
a random realisation of a halo which underlies a mean power spectrum and the recovered
power spectrum from such a halo should be statistically consistent with the mean.

To ensure this feature, we performed a test.  We generated three density fields which are the
random realisation of the same power spectrum ($\propto k^{-3}$ at small scales) as shown in figure
\ref{fig:simgrid}. The three density grids enclose structures and some substructures and look very
different morphologically, but as they are generated from the statistics, they should show
a similar power spectrum. In the bottom-right of figure \ref{fig:simgrid} we show the ratio of the
recovered power spectrum from each of the grids to the input power spectrum. The slope
of the recovered power spectrum is calculated by fitting the power law towards the larger
values of $k$, between 0.68 to 4 (in arbitrary units).
All power spectra give nearly the same slope as the input one with an error of about 10-20$\%$.
This test shows that given a density field, the correct mean statistic can be obtained within
a reasonably small error.

For a pure noise field, we expect same power at all scales. We performed this test as well (but we
don't show the result in a figure form), and verified that the power spectrum is flat.

The fluctuations in mass distribution at various scales can be directly inferred as the
presence/absence of substructures. Larger power at small scale indicates the presence of local
density peaks, however, a flat power at all scales indicates similar structures at all scales
or just the noise.  Also, a smoother distribution of the matter gives rise to smaller power whereas
sharp peaks correspond to larger power. Therefore, we expect the power at small scales to
increase as the merger progresses and reach the highest power when the system is virialised.
In other words, one should expect the power on small scales to decrease with redshift.
This form of the power spectrum can be used in order to compare halos statistically, within
simulations or observations or across the two.


\section{Lens reconstruction method}\label{sec:grale}

In GRALE \citep{2007MNRAS.380.1729L,2008MNRAS.389..415L} the mass
distribution is free-form and consists of a superposition of a large
number of adjustable components (several hundred Plummer lenses).  The
distribution of these Plummer lenses is adaptively determined by GRALE
using a multi-objective genetic algorithm, to optimise two fitness
measures.

\begin{enumerate}

\item The {\em overlap fitness\/} quantifies the fractional overlap of
  the projected images of the same source back onto the source plane.
  If all images of a source back-project to exactly the same area on
  the source plane, the source fitness for that image system is
  perfect.  More generally, the larger overlap the better the fitness.
  It is important to use the fractional overlap, as otherwise the
  fitness measure would be biased towards extreme magnifications.

\item The {\em null fitness\/} is a penalty for any spurious images
  implied by a model where none are present in the observations.
  A null space is created at the image plane, subdivided into a number of
  triangles, and the trial solution
  under study is used to project these triangles onto the source plane. Then, the
  amount of overlap between each triangle and the current estimate of the source
  shape is calculated and used to construct a null space fitness measure which
  is the penalty. The
  penalty is applied only in regions where the data make it clear that
  no images are present; extra images are allowed in regions that are
  difficult to observe because, for example, of the presence of nearby
  bright galaxies.  Among lens-modelling techniques, GRALE is unique
  in being able to exploit the {\em absence\/} of images as useful
  information.

\end{enumerate}

Further fitness measures can be defined and incorporated into GRALE,
in particular, time delays
\citep{2009MNRAS.397..341L,2015PASJ...67...21M}.  No information about
the light from the cluster is used --- mass-traces-light assumptions
are completely absent.

While the genetic algorithm optimizes the properties and placement of
the component Plummer lenses, the user still needs to specify a range
for the allowed number of components.  This is, in effect, an overall
resolution of the mass map.  To set this effective resolution, we run
the mass reconstruction process first at coarse resolution, then
finer, and then coarse again, and choose an optimal trade-off between
fitness and resolution --- see \cite{2014MNRAS.439.2651M} for details
and tests of this strategy.  Finally, the entire cycle is repeated 30
times (while using pre-HFF data) or 40 times (while using post-HFF data),
to generate an ensemble of mass reconstructions. In the remainder
of the article, we present the average mass map of 30 (or 40) mass reconstructions
for each cluster lens, unless stated otherwise. The power spectra
for each of the reconstructions were computed, and both average power spectra
as well as the power spectra of the average mass map are presented. For the true
error estimates, the former is more useful.


\section{Hubble Frontier Fields}\label{sec:hff}

Hubble Frontier Fields (HFF) survey (PI: J.Lotz, HST 13498) is a three year
Director's Discretionary Time program that devotes a total of 840 orbits to six
galaxy clusters plus the accompanying parallel fields.  Each field is observed in
three HST optical bands (ACS F435W, F606W and F814W) and four infra-red bands (WFC3-IR
F105W, F125W, F140W, and F160W).
It is the single most ambitious commitment of HST resources to the exploration of
the distant Universe through the power of gravitational lensing by massive galaxy
clusters. All six clusters are early or intermediate stage mergers at $z=0.3$--0.55,
with significant elongation and hence a non-trivial mass distribution. Each had about
10--20 multiply imaged background sources discovered with pre-HFF HST
data. Five independent teams were tasked with making mass and magnification maps for
these six clusters. In the analysis of the present paper, we use mass maps presented
by our team, made with pre-HFF lensing data on six clusters as well as two maps made
with post-HFF data, for clusters MACS0416 and Abell 2744. Table \ref{tbl:HFF} shows the necessary
information about the lensing data in each of the HFF clusters.
In the following figures, the mass maps and the corresponding power spectra are
marked as $\mathtt{v1}$ and $\mathtt{v3}$ indicating pre-HFF and post-HFF data respectively.

\begin{table*}
\centering
\begin{tabular}{ l | l | c | c | c | c  }
  \hline
  Cluster & Data & Redshift & Number of images & Number of sources & Redshift range of sources \\
  \hline
  MACS-J0416 & pre-HFF & 0.396 & 40 & 13 & 1.82-3.25 \\
  MACS-J0416 & post-HFF & 0.396 & 88 & 36 & 1.00-5.90 \\
  Abell 2744 & pre-HFF & 0.308 & 41 & 12 & 2.00-4.00 \\
  Abell 2744 & post-HFF & 0.308 & 55 & 18 & 2.00-4.00 \\
  Abell 370 & pre-HFF & 0.375 & 36 & 11 & 0.73-3.00 \\
  AS-1063 & pre-HFF & 0.348 & 37 & 13 & 1.22-6.09 \\
  MACS-1149 & pre-HFF & 0.543 & 32 & 11 & 1.23-3.80 \\
  MACS-0717 & pre-HFF & 0.545 &  23 & 7 & 1.80-2.91 \\
  \hline
\end{tabular}
\caption{Lensing data from different HFF clusters.}
\label{tbl:HFF}
\end{table*}

\subsection{MACS J0416.1-2403}\label{sec:0416}
\begin{figure*}
	\centering
	\includegraphics[width=0.48\textwidth]{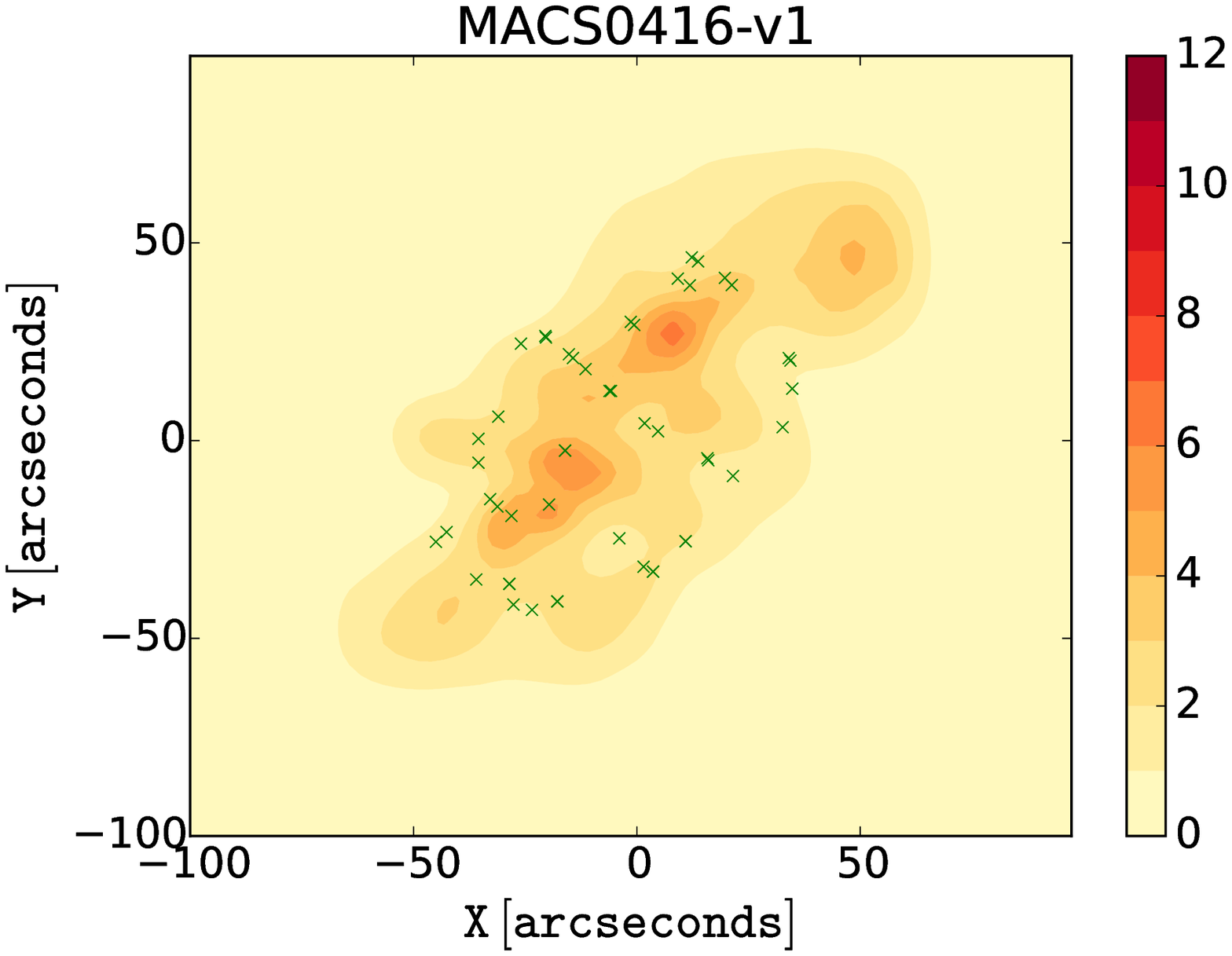}
	\includegraphics[width=0.48\textwidth]{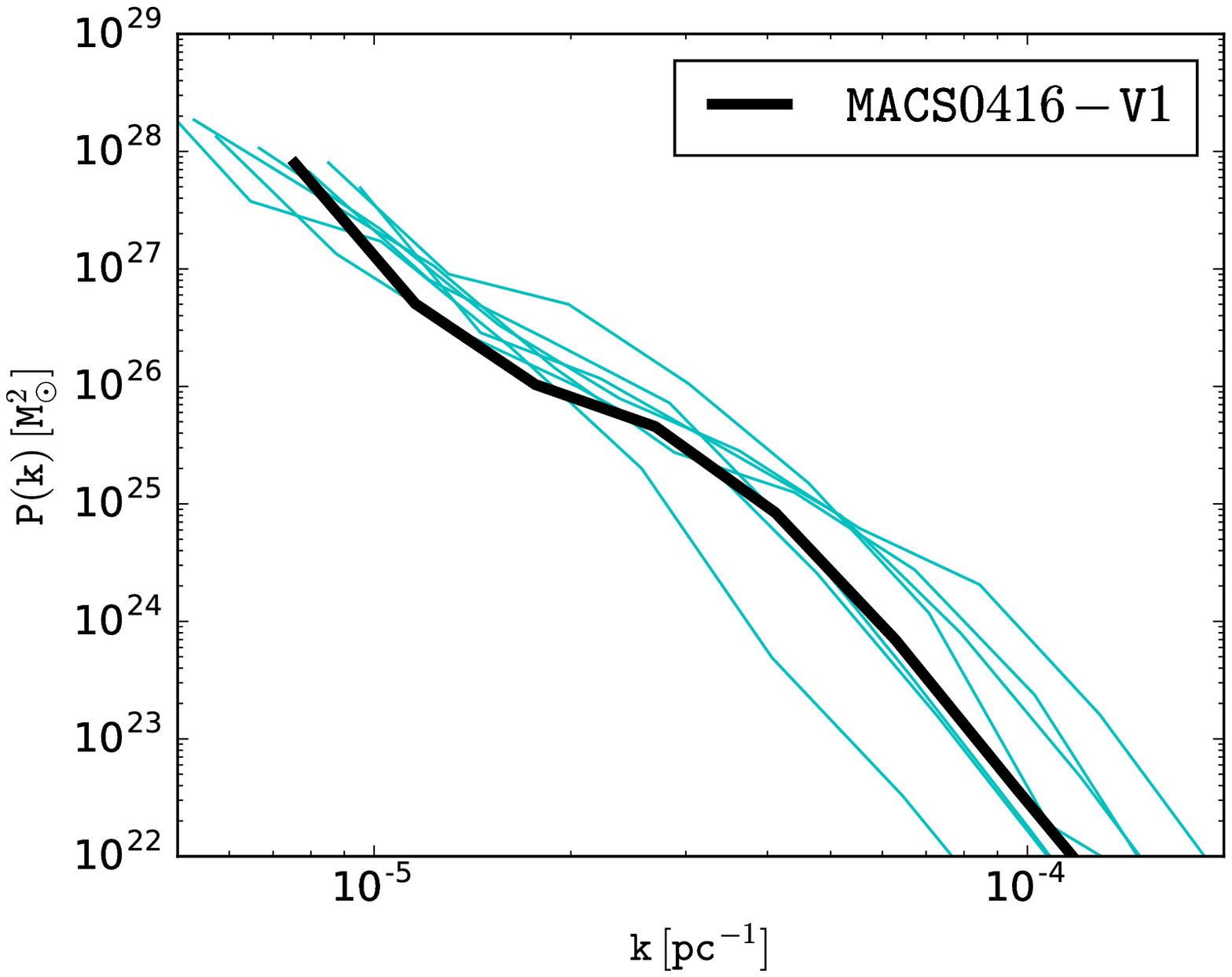}\\
	\includegraphics[width=0.48\textwidth]{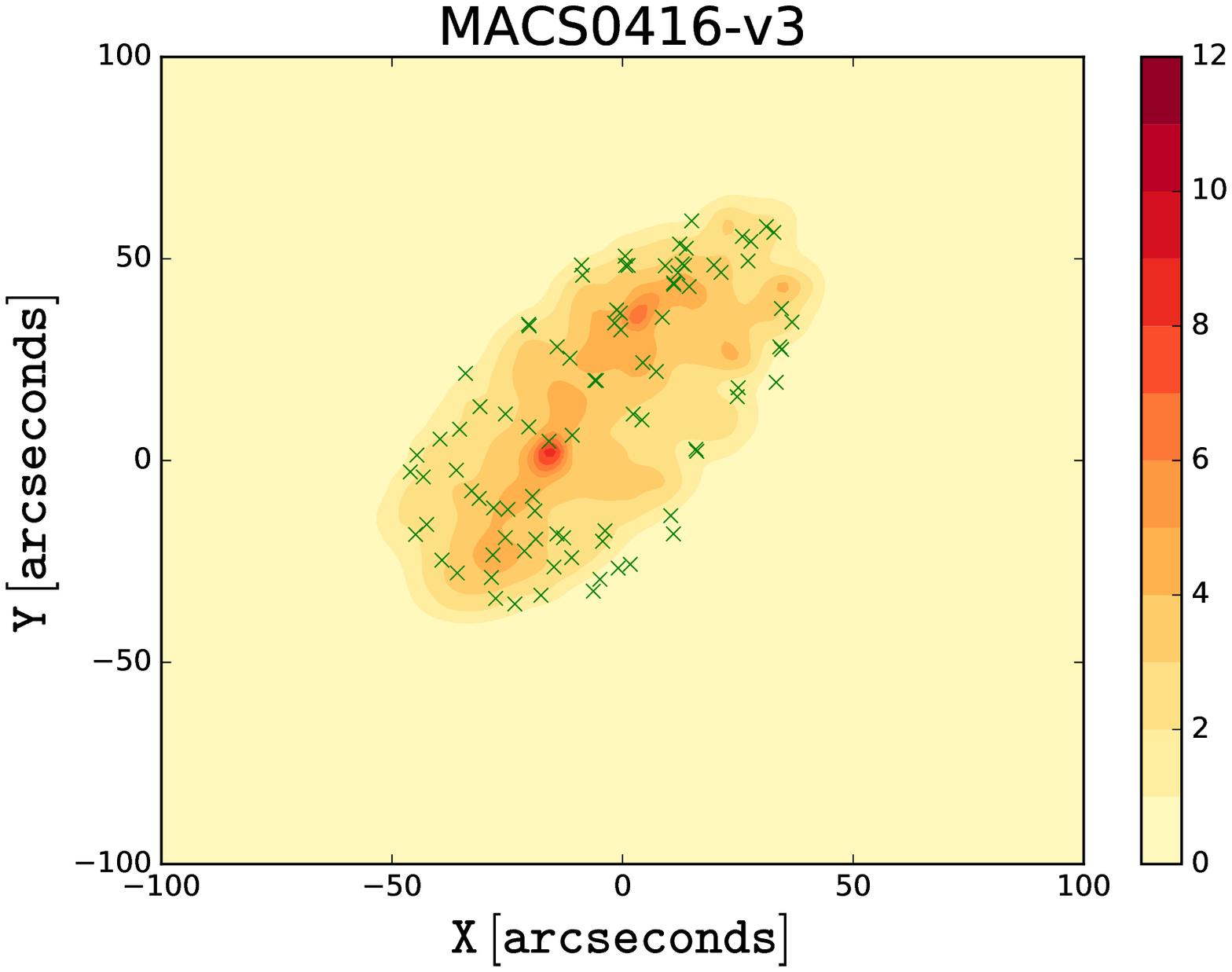}
	\includegraphics[width=0.48\textwidth]{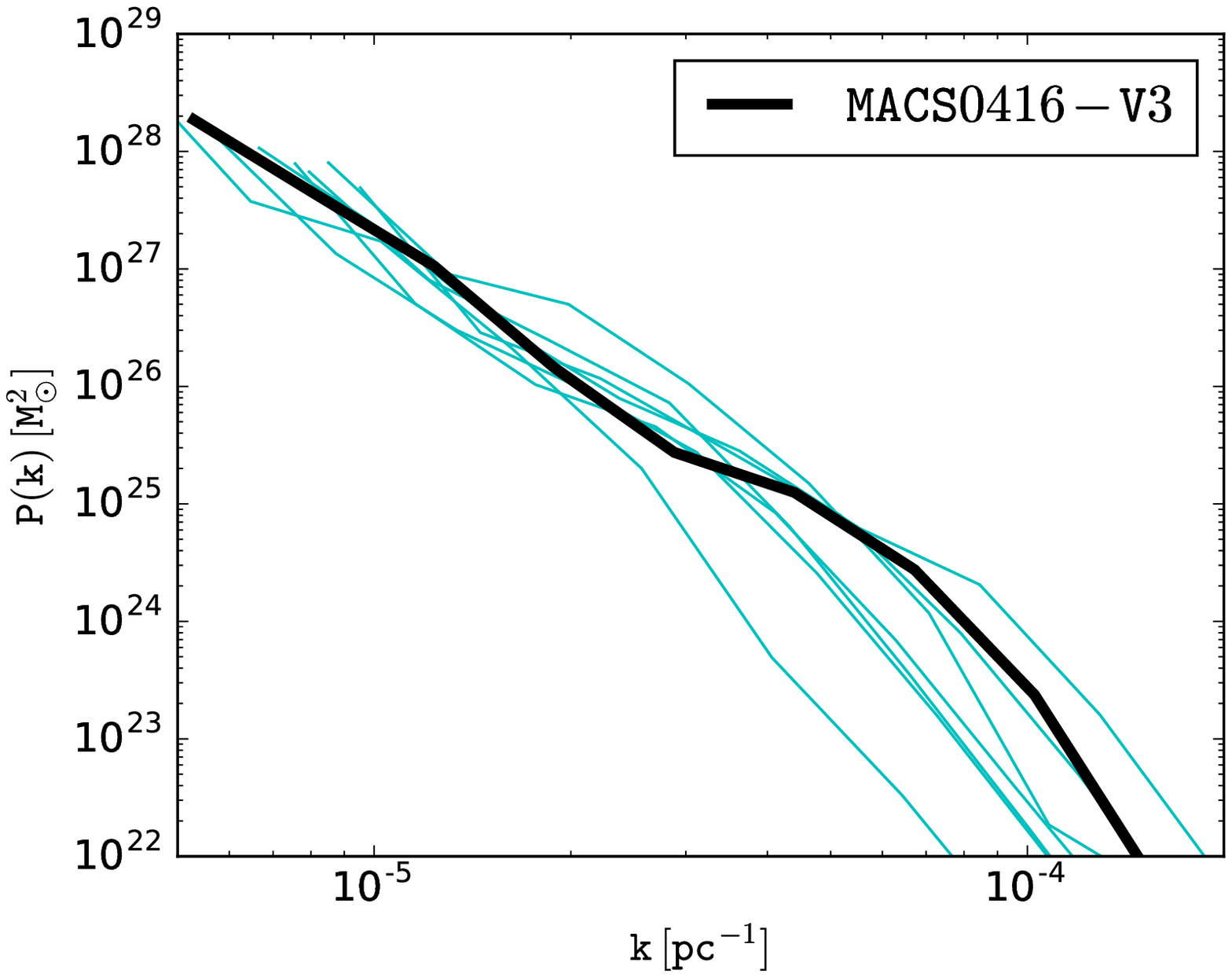}
	\caption{Left: the density map for MACS0416 cluster using pre-HFF
	data (top row) and HFF data (bottom row). The crosses mark the positions
	of the lensed images.
	Right: the highlighted (black)
	corresponding power spectrum of the cluster on the left with all other
	HFF clusters' power spectra in the background.}
	\label{fig:density:0416}
\end{figure*}

MACS J0416.1-2403 (MACS0416 henceforth) is a strong gravitational lens at redshift of
nearly 0.396. It is an elongated galaxy cluster, most probably a recent merger or a pre-merger
\citep{ogr15}, and hence has a non-trivial mass distribution. It was first identified by the MAssive
Cluster Survey \citep[MACS;][]{2010MNRAS.407...83E}. Based on its double-
peaked surface brightness in the X-rays \citep{2012MNRAS.420.2120M},
its recent merger stage is confirmed. It is amongst the five high magnification
clusters in the CLASH (Cluster Lensing and Supernovae survey with Hubble) project
\citep{2012ApJS..199...25P}. Detailed mass maps were provided by
\cite{2013ApJ...762L..30Z} using CLASH data, and \cite{2015MNRAS.446.4132J}
and \cite{2015arXiv150708960S} using HFF data.

We made two reconstructions of this lens using GRALE, one with pre-HFF data (total
40 images from 13 sources), using data from \cite{2011MNRAS.417..333M}, as well as data provided by
J. Richard and D. Coe, and one with HFF data (total 88 images from 36 sources) 
where we used images that were classified collectively by the HFF teams
being of GOLD and SILVER quality. Many of these images were initially identified by
\cite{2015MNRAS.446.4132J}. The two reconstructed mass models, $\mathtt{v1}$ and
$\mathtt{v3}$---with each being an average
of 30 and 40 realisations respectively---are shown in figure \ref{fig:density:0416} (left column),
each with its respective power spectrum (highlighted in the right column). Both mass models
are very similar at large scales, which is also evident from the power spectrum, and
only differ in the details at small scales. The two mass clumps and the elongation
can be identified in both models. However, the mass models with HFF data (bottom row),
shows larger power or many detailed substructures at small scales. This is
expected as HFF data has many more lensed images than pre-HFF data and hence provides additional
lensing constraints at small scales, which leads to increased power at larger $k$'s.

\subsection{Abell 2744}\label{sec:2744}
\begin{figure*}
	\centering
	\includegraphics[width=0.48\textwidth]{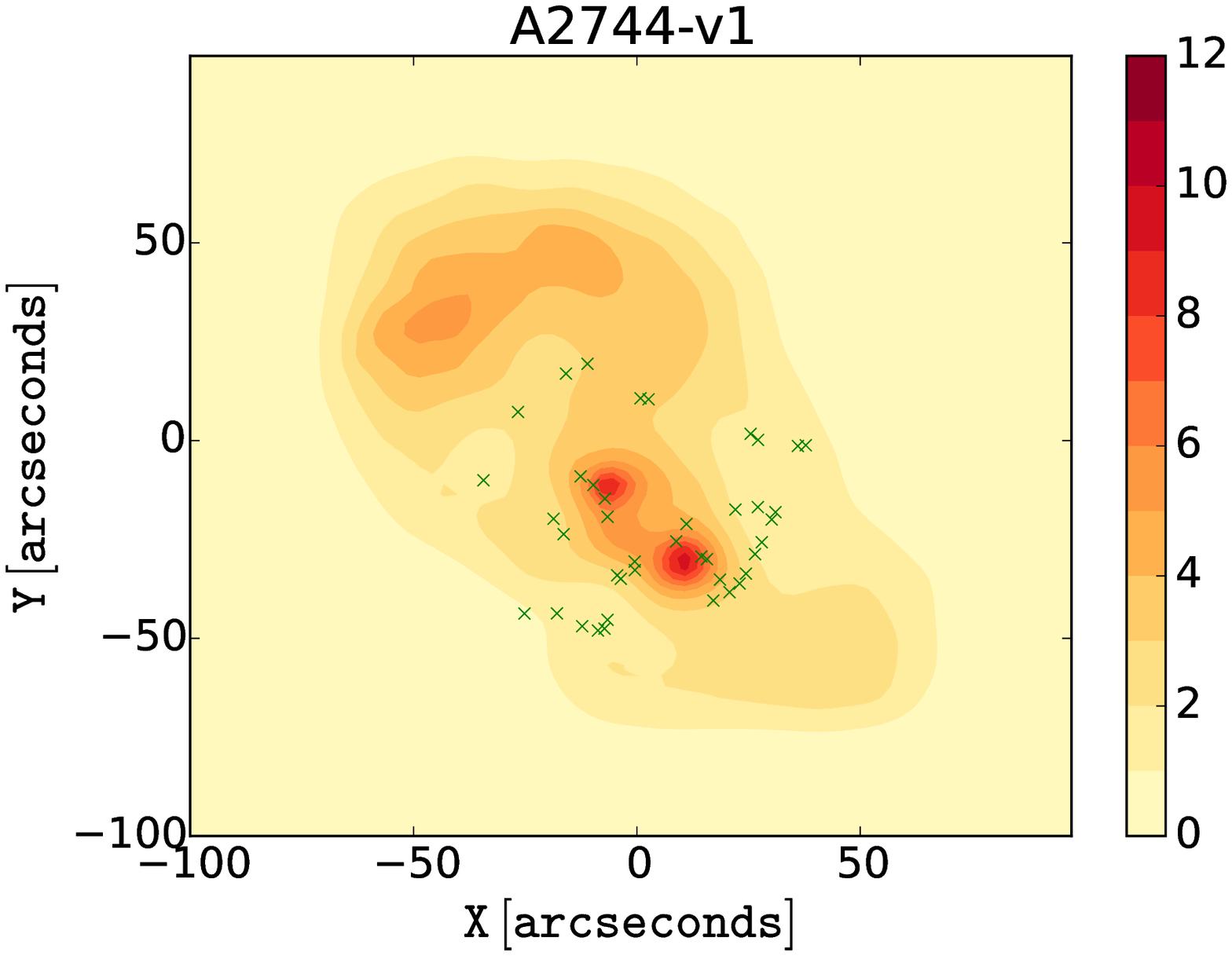}
	\includegraphics[width=0.48\textwidth]{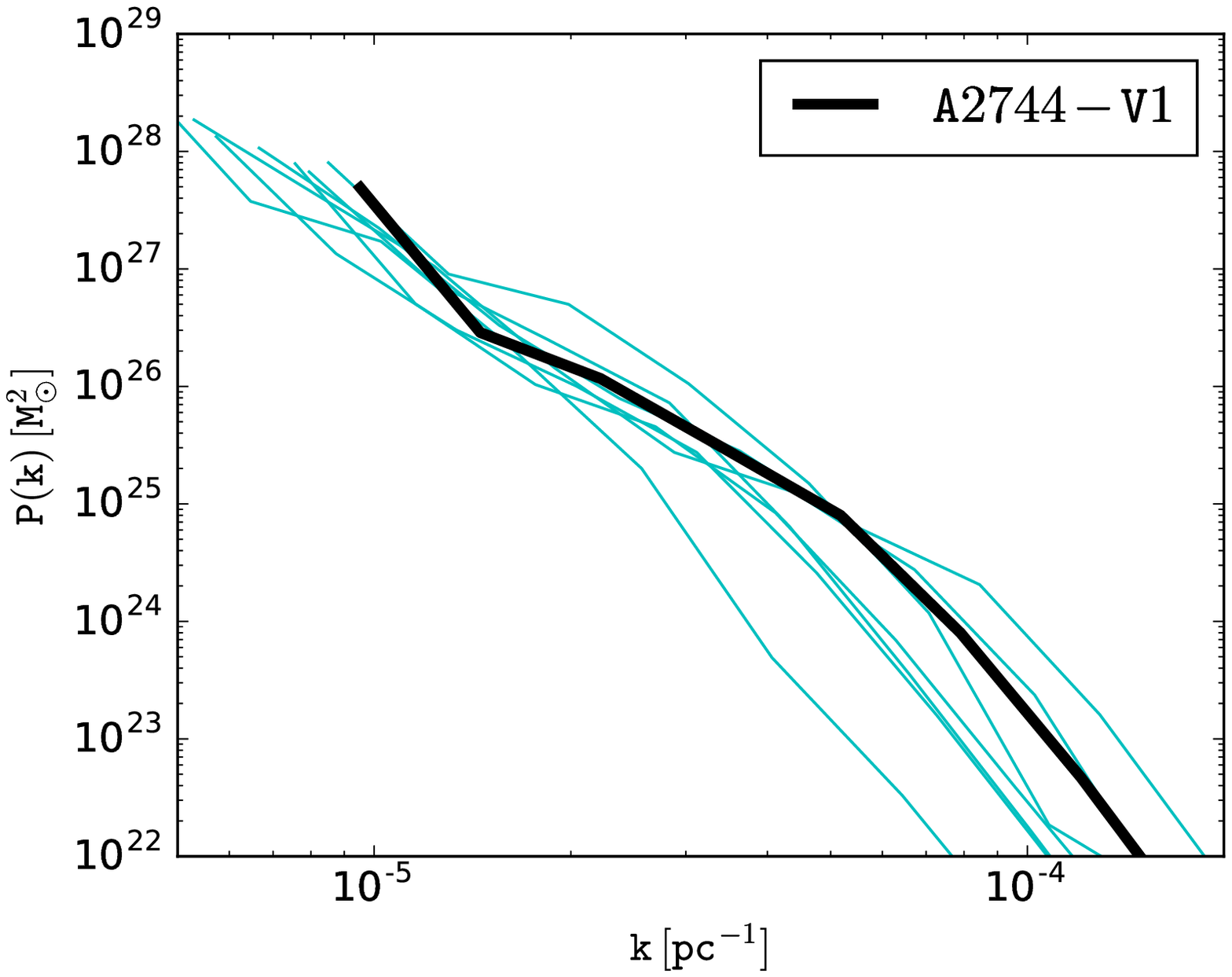}
  \includegraphics[width=0.48\textwidth]{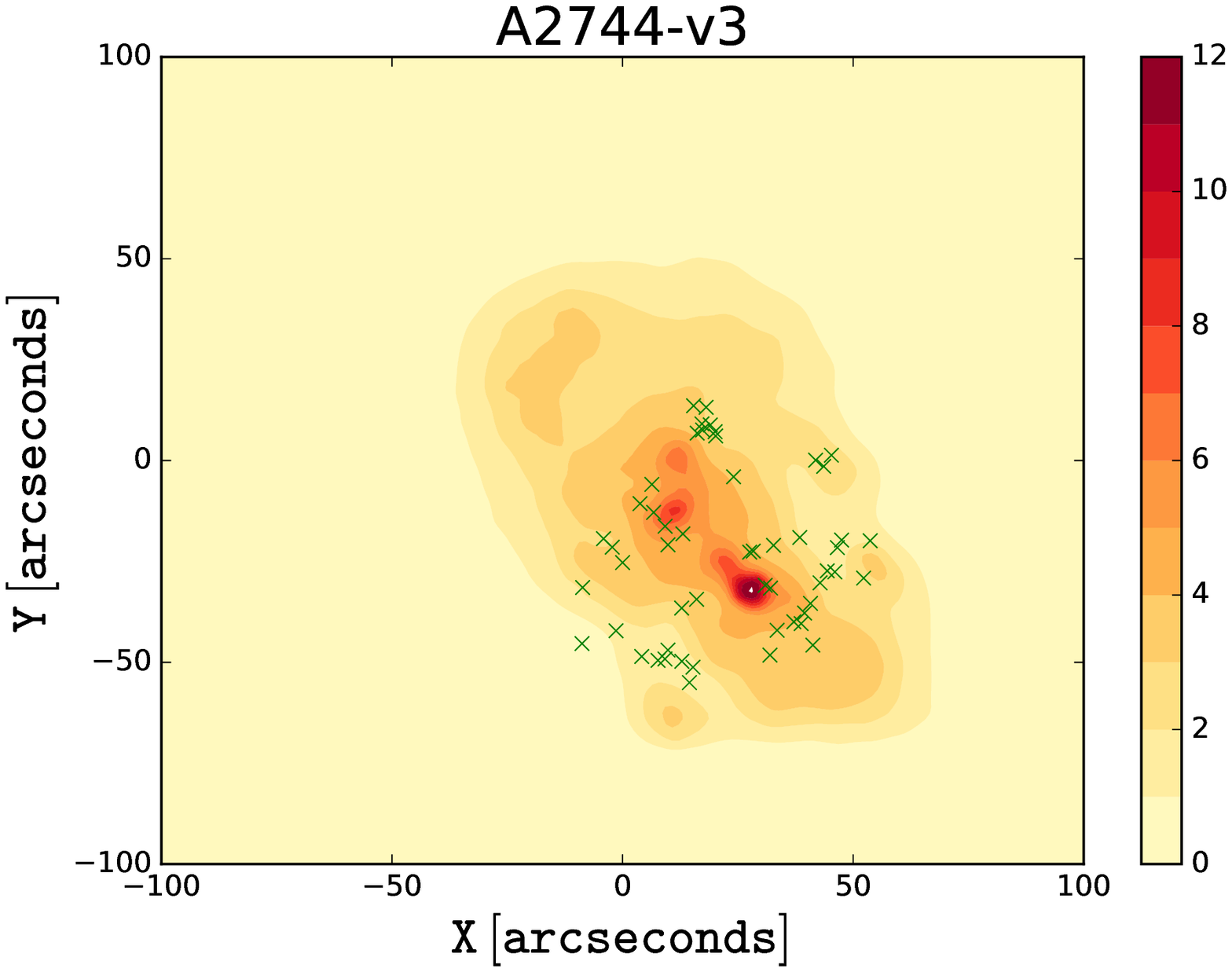}
  \includegraphics[width=0.48\textwidth]{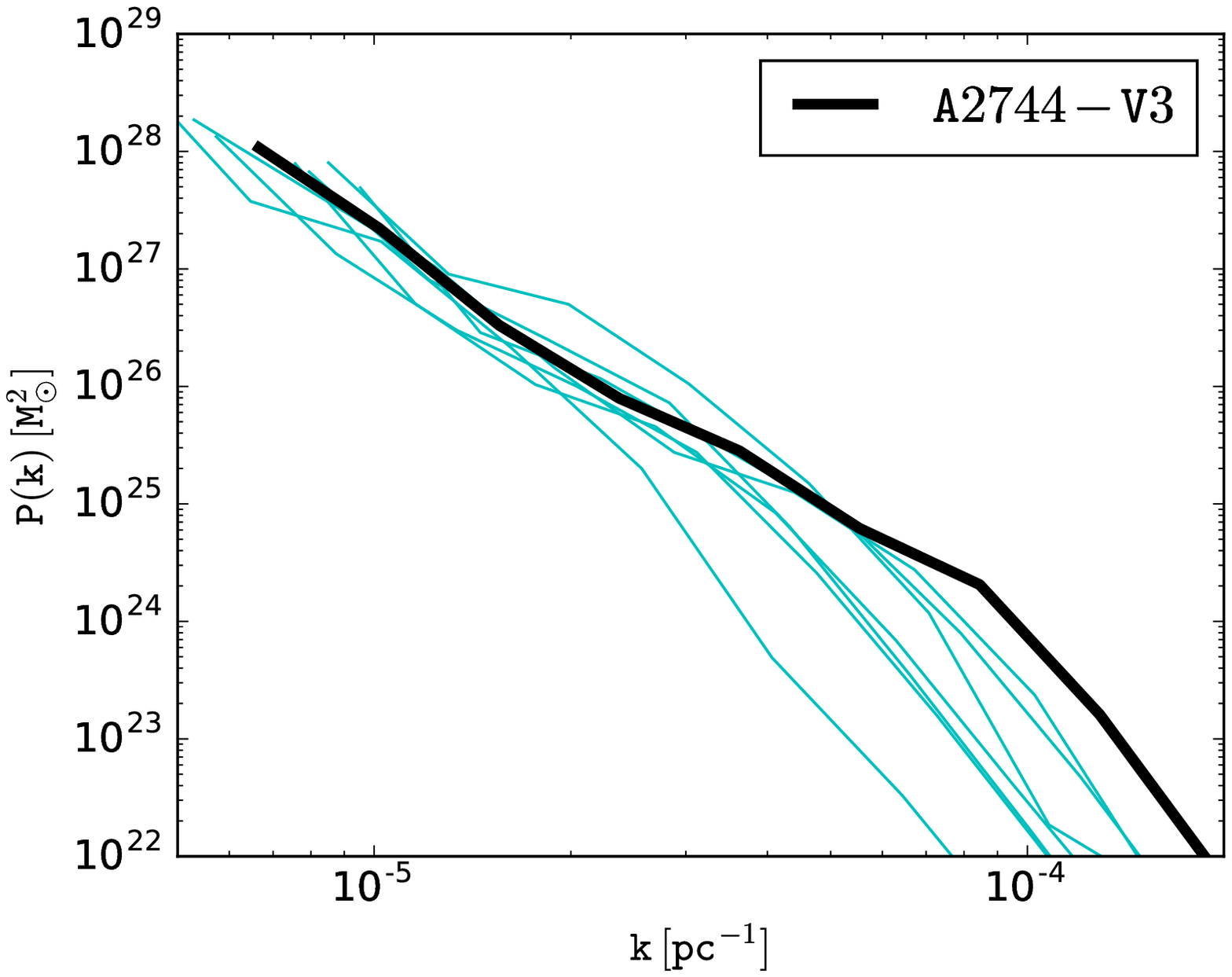}
	\caption{Left: the density map for the Abell 2744 cluster using
	 pre-HFF data (top row) and post-HFF data (bottom row). 
   The upper left clump could well be an
	 artefact as there
	 are no images there.
	 Right: the highlighted (black) corresponding power
	 spectrum of the cluster on the left with all other
	HFF clusters' power spectra in the background.}
	\label{fig:density:2744}
\end{figure*}

Abell 2744 is a massive galaxy cluster at redshift 0.308 and an active merger.
It has been studied in various wavelengths, for example, it has an extended radio halo
\citep{1999NewA....4..141G,2001A&A...369..441G}, X-ray emission
\citep{1998MNRAS.296..392A,2010MNRAS.407...83E,2004MNRAS.349..385K,2011ApJ...728...27O}
and various substructures have been identified in the optical observations
\citep{2001ApJ...548...79G,2006A&A...449..461B}.
\cite{2010MNRAS.406.1134S} show a significant offset between the dark-matter
component (from lensing) and baryonic matter (from X-rays observations).
There is also a magnified singly imaged supernova in the background,
at redshift 1.35 \citep{rod15}.
Various lensing analyses have been done
\citep{1997ApJ...479...70S,1998MNRAS.296..392A,2004ApJ...613...95C,
2011MNRAS.417..333M,2014ApJ...797...98L,2014arXiv1409.8663J}.

We present two mass models of Abell 2744 using GRALE, one with pre-HFF data
\citep{2011MNRAS.417..333M,2014MNRAS.444..268R} as well as data provided by J. Richard and D. Coe, for
 a total of 41 images from 12 background sources spread over a redshift range of 2 to 4 and
 another with post-HFF data for a total of 55 images from 18 sources. Figure
\ref{fig:density:2744} shows the reconstructed mass maps and the corresponding
power spectra. It shows two very distinct blobs and an overall
elongated structure, a morphology very likely for a major merger which is in fact confirmed
by previous studies. The post-HFF data contains improved identifications and positions
of the lensing images, and the corresponding map (bottom row of figure \ref{fig:density:2744}) shows fewer spurious structures compared to the pre-HFF map (top row of figure \ref{fig:density:2744}). 

Due to very sharp mass peaks, this cluster, like MACS0416, also shows larger power at small scales.

\subsection{Abell 370}\label{sec:370}
\begin{figure*}
	\centering
	\includegraphics[width=0.48\textwidth]{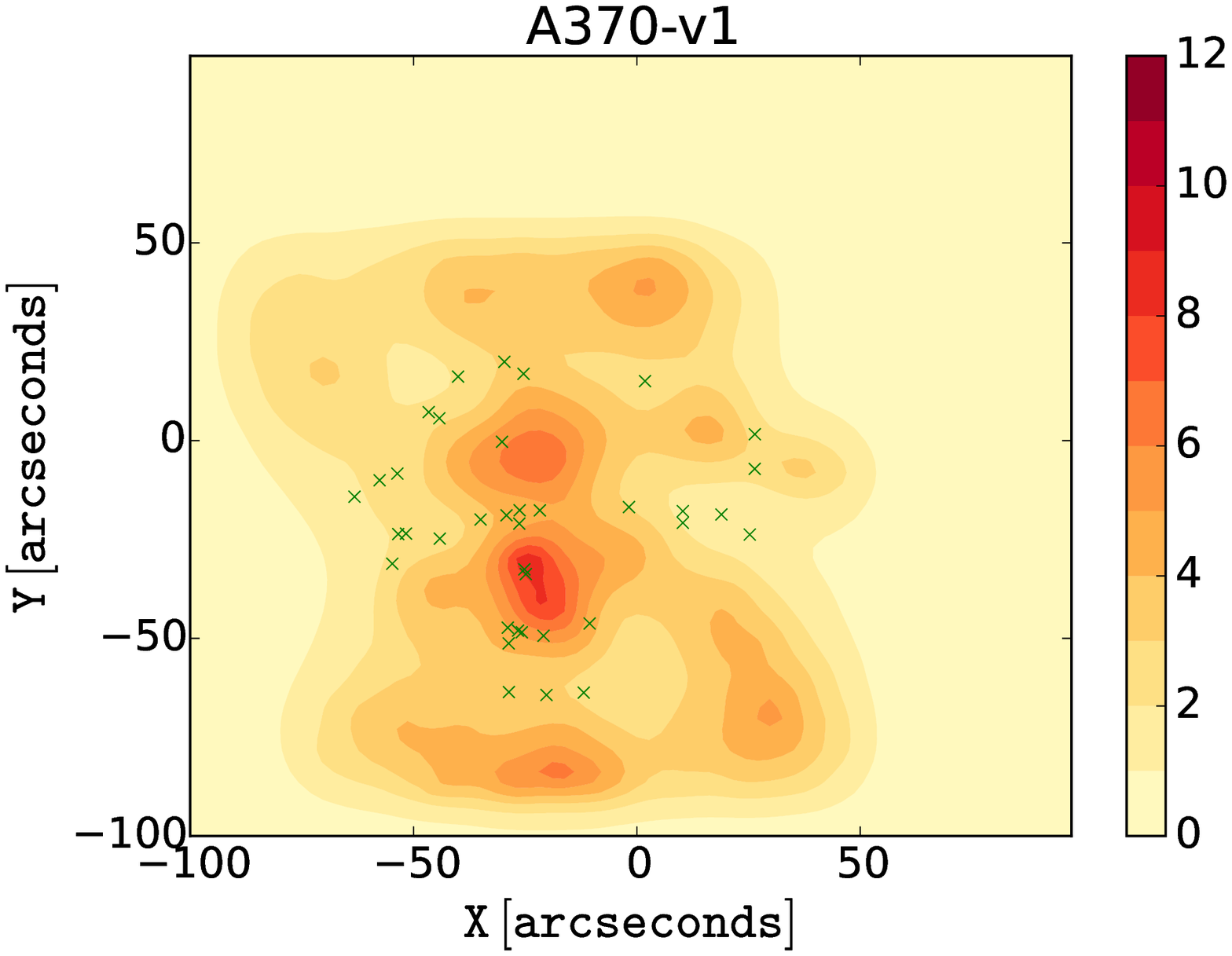}
	\includegraphics[width=0.48\textwidth]{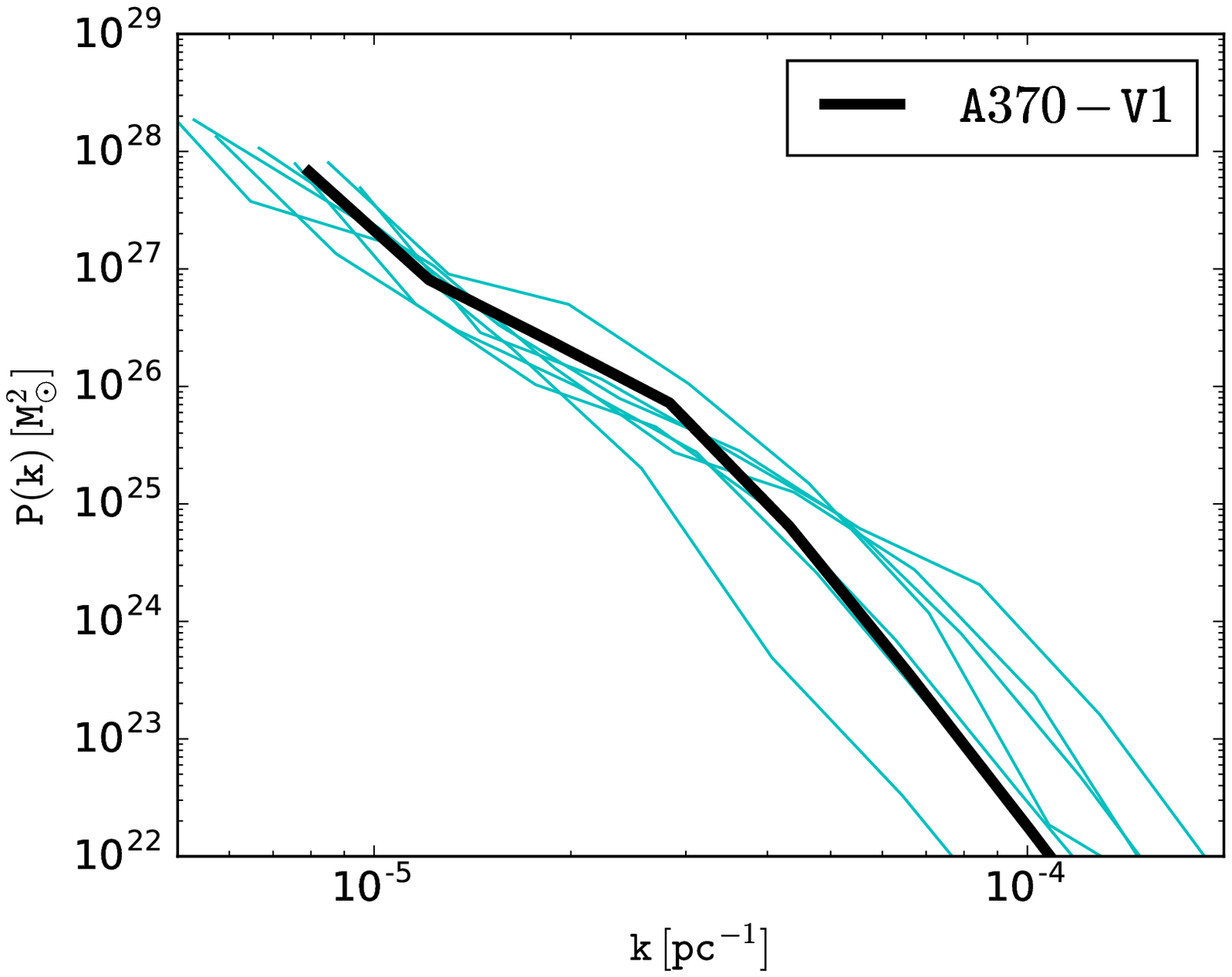}
	\caption{Left: the density map for the Abell 370 cluster using
	 pre-HFF data. Right: the highlighted (black) corresponding power
	 spectrum of the cluster on the left with all other
	HFF clusters' power spectra in the background.}
	\label{fig:density:370}
\end{figure*}

Abell 370 is a strong gravitational lens at redshift 0.375 and hosts a giant
gravitational arc at redshift 0.725. Due to its large Einstein radius, it is
an ideal target to look for high redshift galaxies through its magnification, especially
in high-density regions. There are many published mass models
\citep[for example][]{1987Msngr..50....5S,1998MNRAS.294..734A,ric10}.
We used pre-HFF data from the latter work, as well as data provided by D. Coe and A. Koekemoer,
to reconstruct its mass distribution. Figure \ref{fig:density:370}
shows the resulting mass map. The mass distribution is shallow but shows many
substructures. The resulting power spectrum shows less power on small scales
as compared to other HFF clusters which is reflected in shallower peaks in
its mass distribution.

\subsection{Abell S1063}\label{sec:1063}
\begin{figure*}
	\centering
	\includegraphics[width=0.48\textwidth]{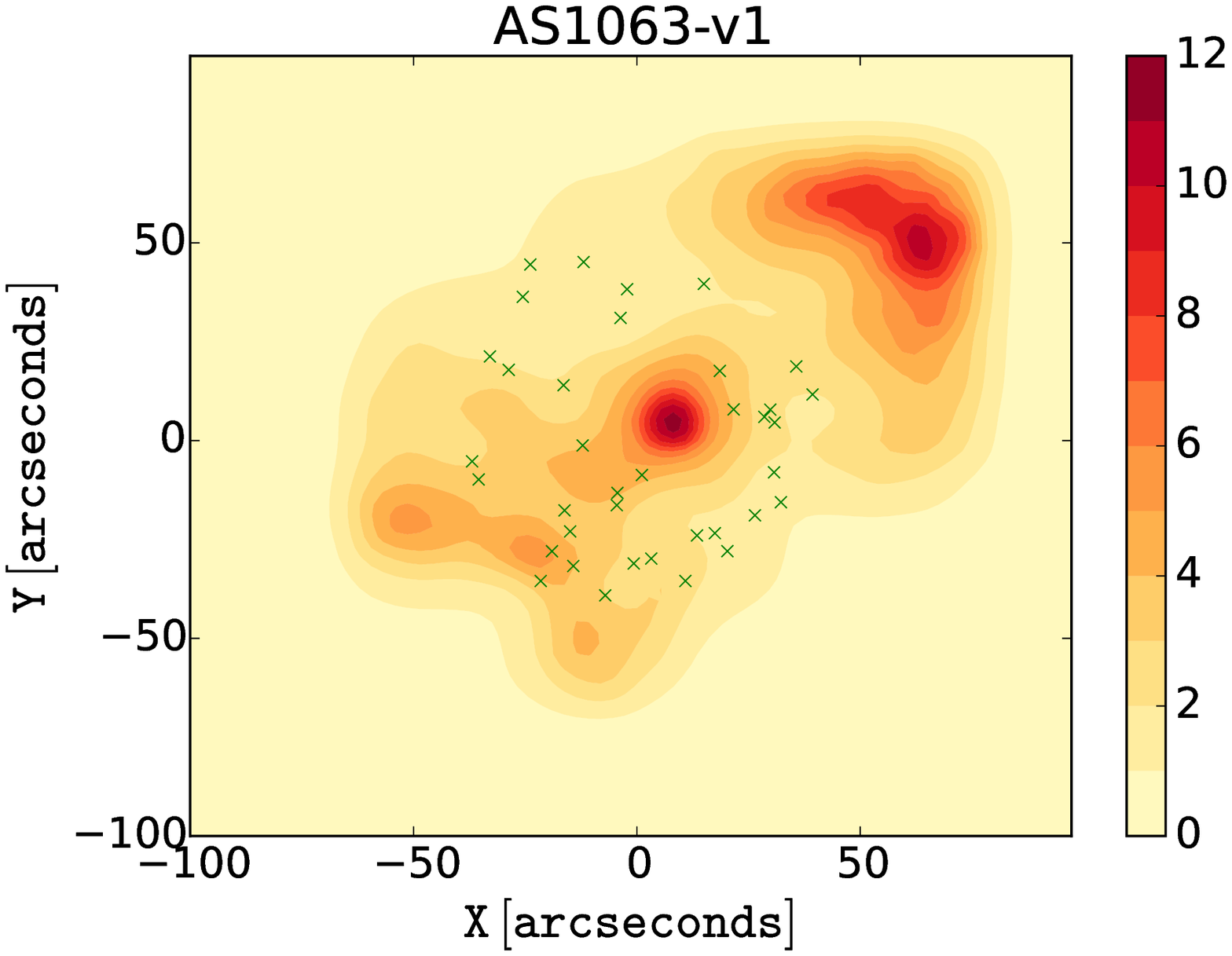}
	\includegraphics[width=0.48\textwidth]{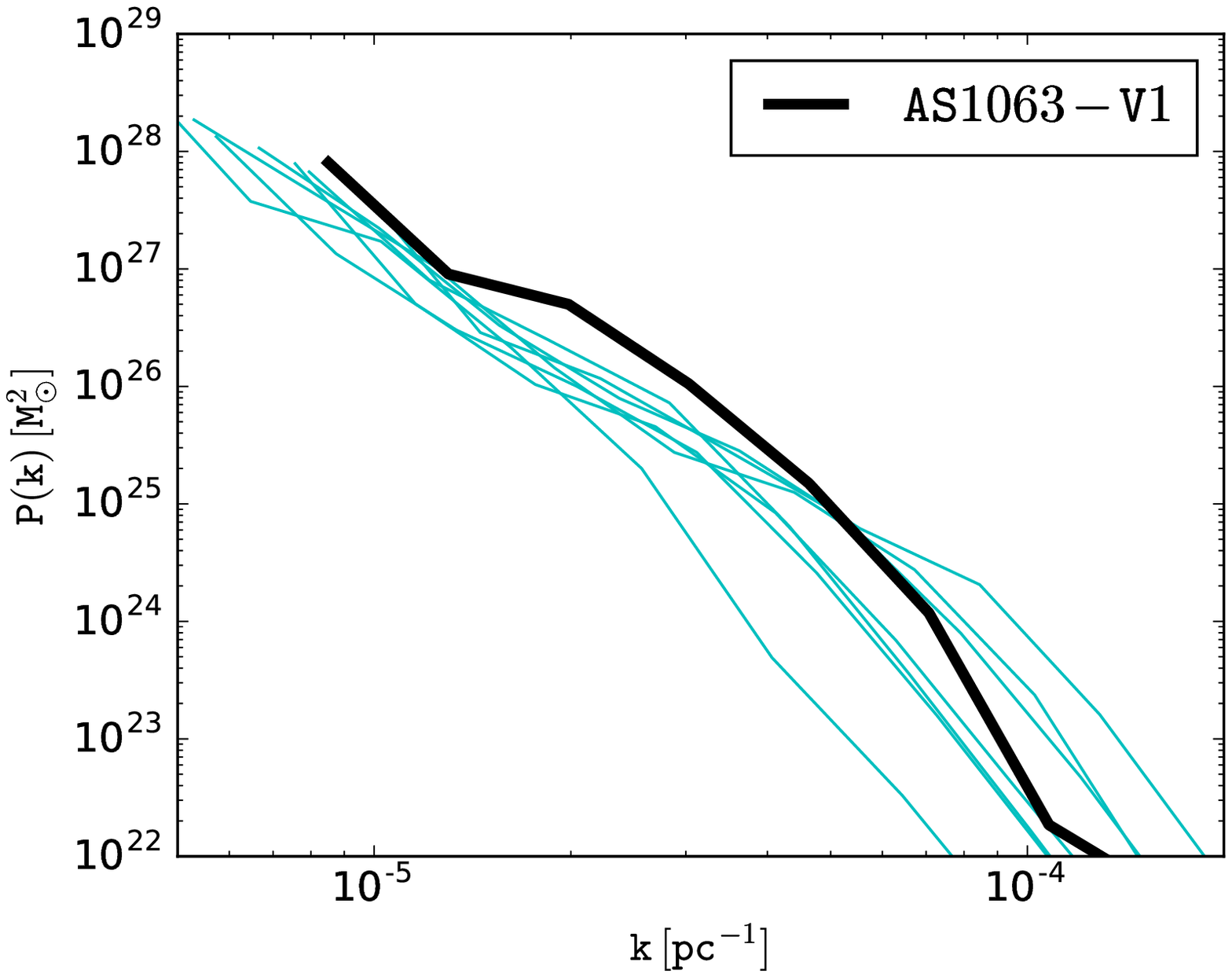}
	\caption{Left: the density map for the AS1063 cluster using
	 pre-HFF data. Right: the highlighted (black) corresponding power
	 spectrum of the cluster on the left with all other
	HFF clusters' power spectra in the background.
	Note that there are no images in the upper right corner of the map, so the massive
	clump in that region could well be an artefact.
	}
	\label{fig:density:1063}
\end{figure*}

Abell S1063 is a galaxy cluster at redshift 0.348. We used pre-HFF data identified
by \cite{2014MNRAS.444..268R}; our reconstruction used a total of 37 images from 13 background
sources.

Figure \ref{fig:density:1063} shows the reconstructed mass map along with the
corresponding power spectrum. It shows two mass clumps, which lead to
larger power at intermediate scales, but due to the shallowness of the peaks, it loses
power at small scales.

\subsection{MACS J1149.5+2223}\label{sec:1149}
\begin{figure*}
	\centering
	\includegraphics[width=0.48\textwidth]{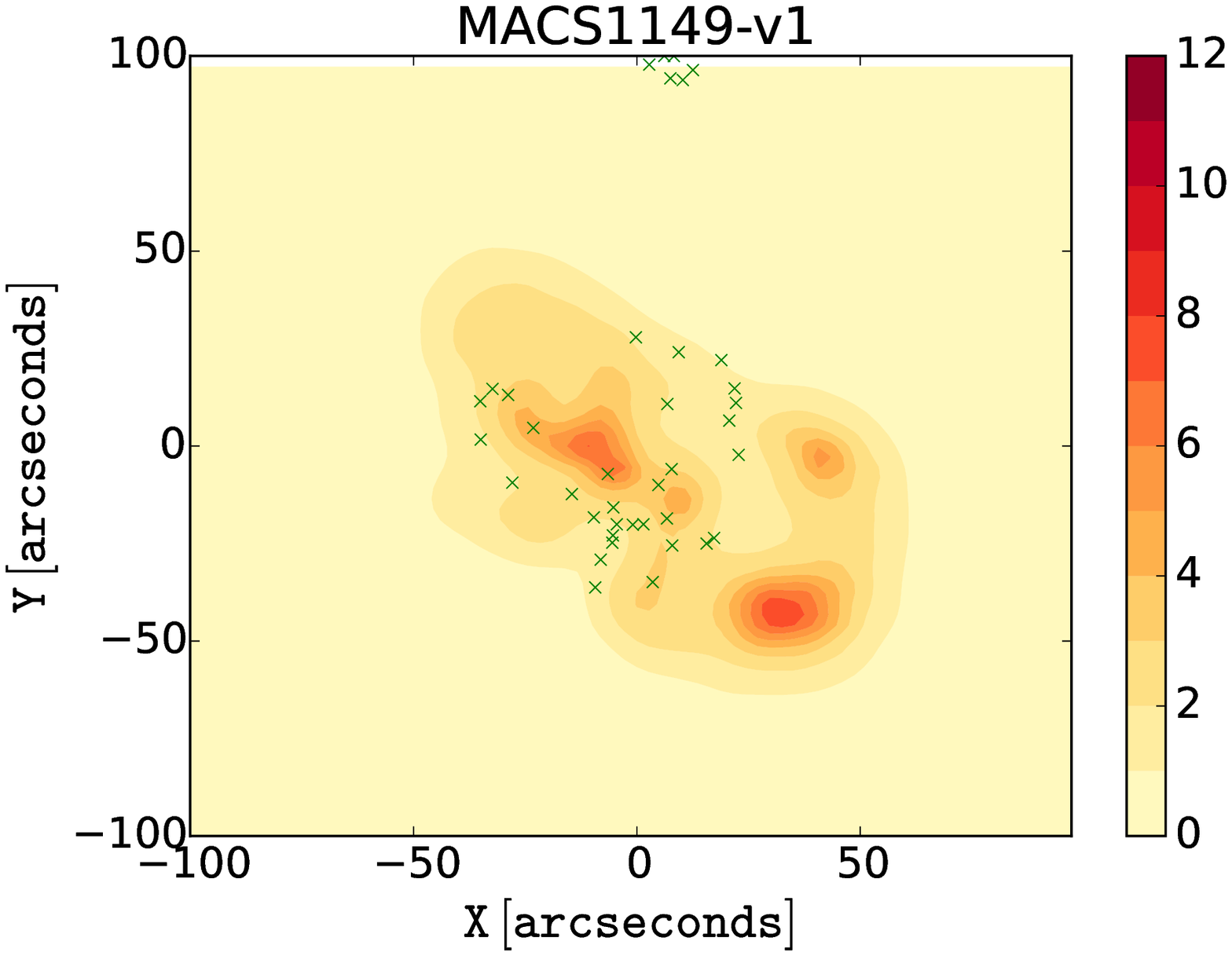}
	\includegraphics[width=0.48\textwidth]{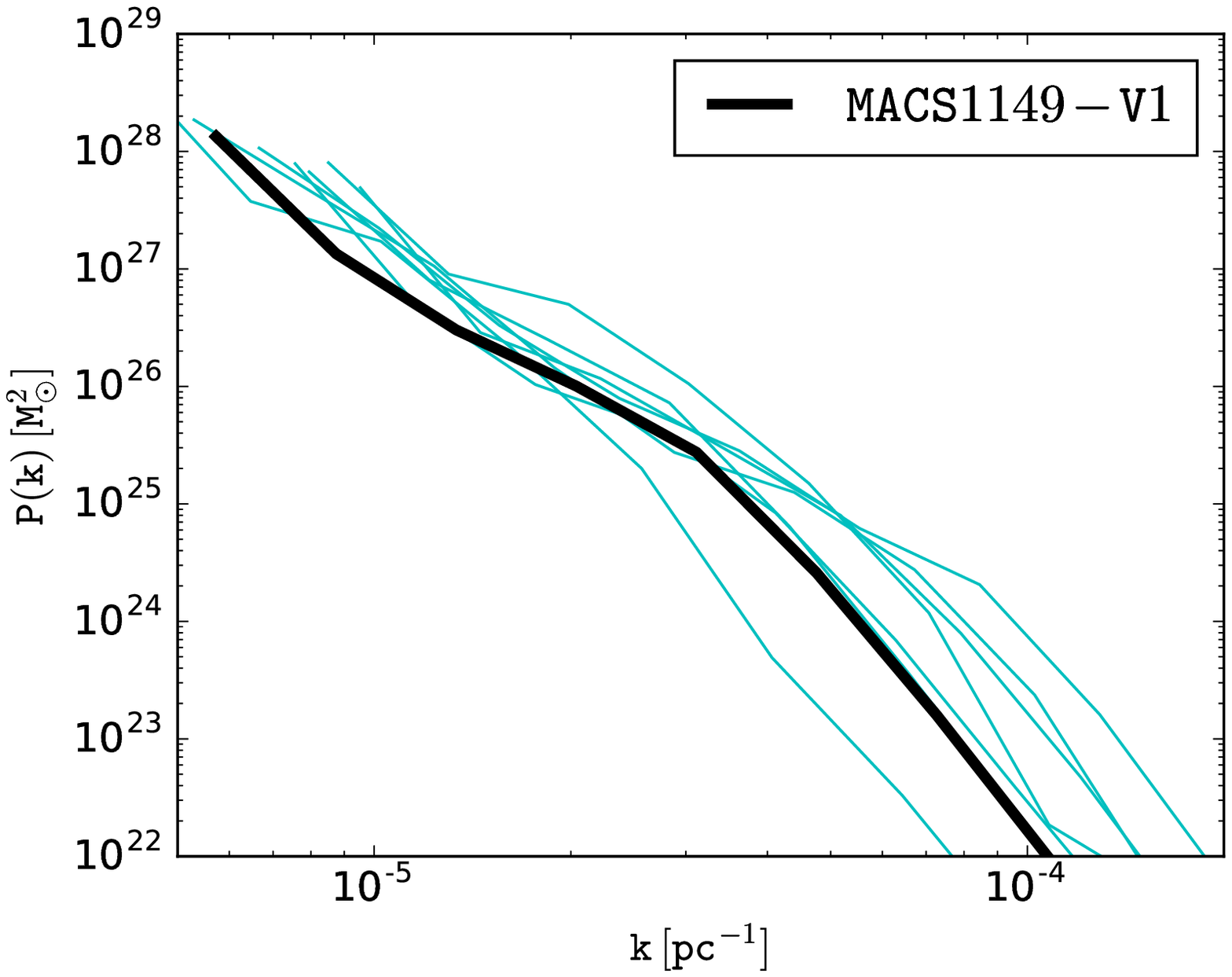}
	\caption{Left: the density map for the MACS1149 cluster using
	 pre-HFF data. Right: the highlighted (black) corresponding power
	 spectrum of the cluster on the left with all other
	HFF clusters' power spectra in the background.
There are no images beyond X=25'', so the two clumps on the right
could well be artefacts.
        }
	\label{fig:density:1149}
\end{figure*}

MACS J1149+2223 (MACS1149 hereafter) is an X-ray bright cluster at redshift 0.543.
It has been studied by various authors for its rich strong lensing
\citep{2009ApJ...707L.163S,2009ApJ...703L.132Z,2011MNRAS.410.1939Z,2015ApJ...801...44Z,2014MNRAS.443..957R,2014ApJ...797...48J,2015ApJ...800L..26S}.
There is also a multiply imaged supernova
\citep{2014ATel.6729....1K,2015Sci...347.1123K} hosted by a face-on spiral galaxy
at redshift 1.49.

Figure \ref{fig:density:1149} shows the reconstructed mass model of MACS1149 using
GRALE with pre-HFF data (total 32 images from 11 background sources) from
\cite{2009ApJ...707L.163S,2009ApJ...703L.132Z}, and data provided by M. Limousin.
The mass
model favours two dominant peaks and one sub-peak, elongation and many substructures.
The peaks are shallower than those in Abell 2744, also expected as it has nearly the
same mass as Abell 2744 but higher redshift. The corresponding power spectrum is shown
in the right column of the same figure.


\subsection{MACS J0717.5+3745}\label{sec:0717}
\begin{figure*}
	\centering
	\includegraphics[width=0.48\textwidth]{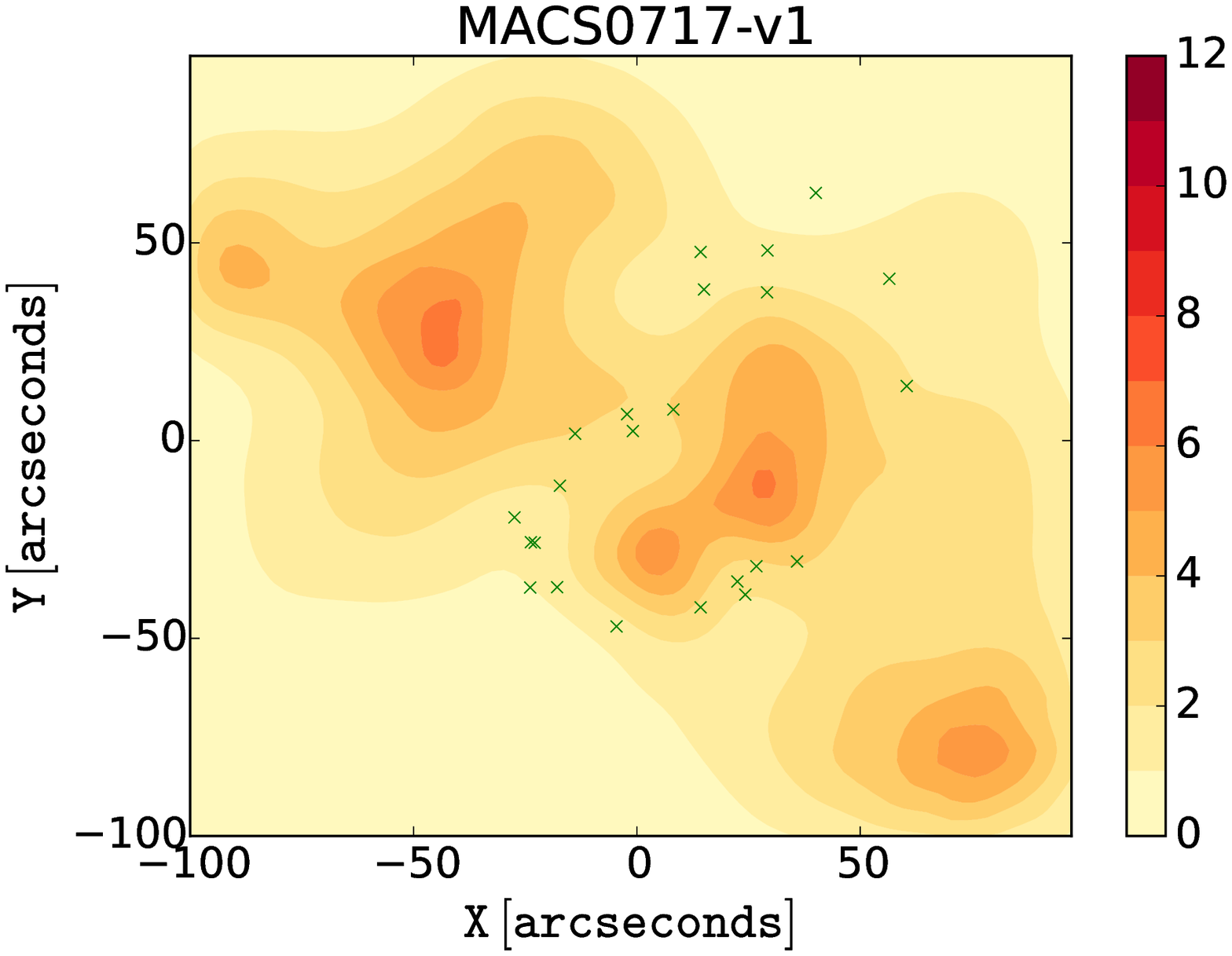}
	\includegraphics[width=0.48\textwidth]{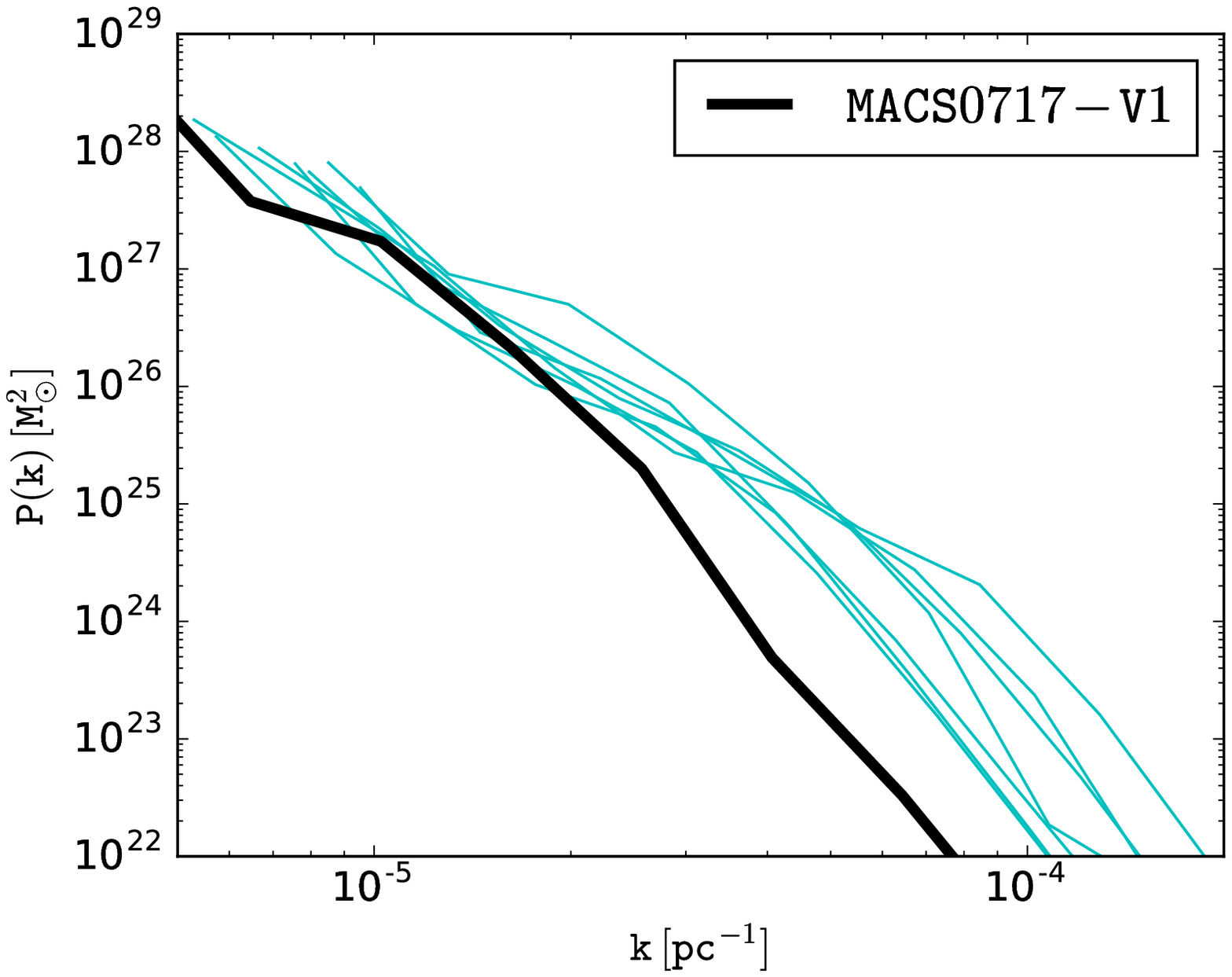}
	\caption{Left: the density map for the MACS0717 cluster using
	 pre-HFF data. The upper left and bottom right clumps could
	 well be artefacts as there
	 are no images there.
	 Right: the highlighted (black) corresponding power
	 spectrum of the cluster on the left with all other
	HFF clusters' power spectra in the background.}
	\label{fig:density:0717}
\end{figure*}

MACS J0717.5+3745 (MACS0717 hereafter) is a strong gravitational lens at redshift 0.55,
classified as the most dramatic merger in X-ray/optical analysis by
\cite{2012MNRAS.420.2120M}.

We used pre-HFF data from \cite{2012A&A...544A..71L,2014MNRAS.444..268R} and \cite{med13} to reconstruct
its mass distribution using GRALE. Figure \ref{fig:density:0717} shows the resulting
mass map and the corresponding power spectrum. Due to the clear mass peaks and substructures,
the power spectrum shows increased power at intermediate scales, however, due to very shallow peaks the
power spectrum at small scales drops and is amongst the lowest of the HFF clusters.


\section{Comparing the clusters}\label{sec:comparison}

\begin{figure*}
	\centering
  \includegraphics[width=0.48\textwidth]{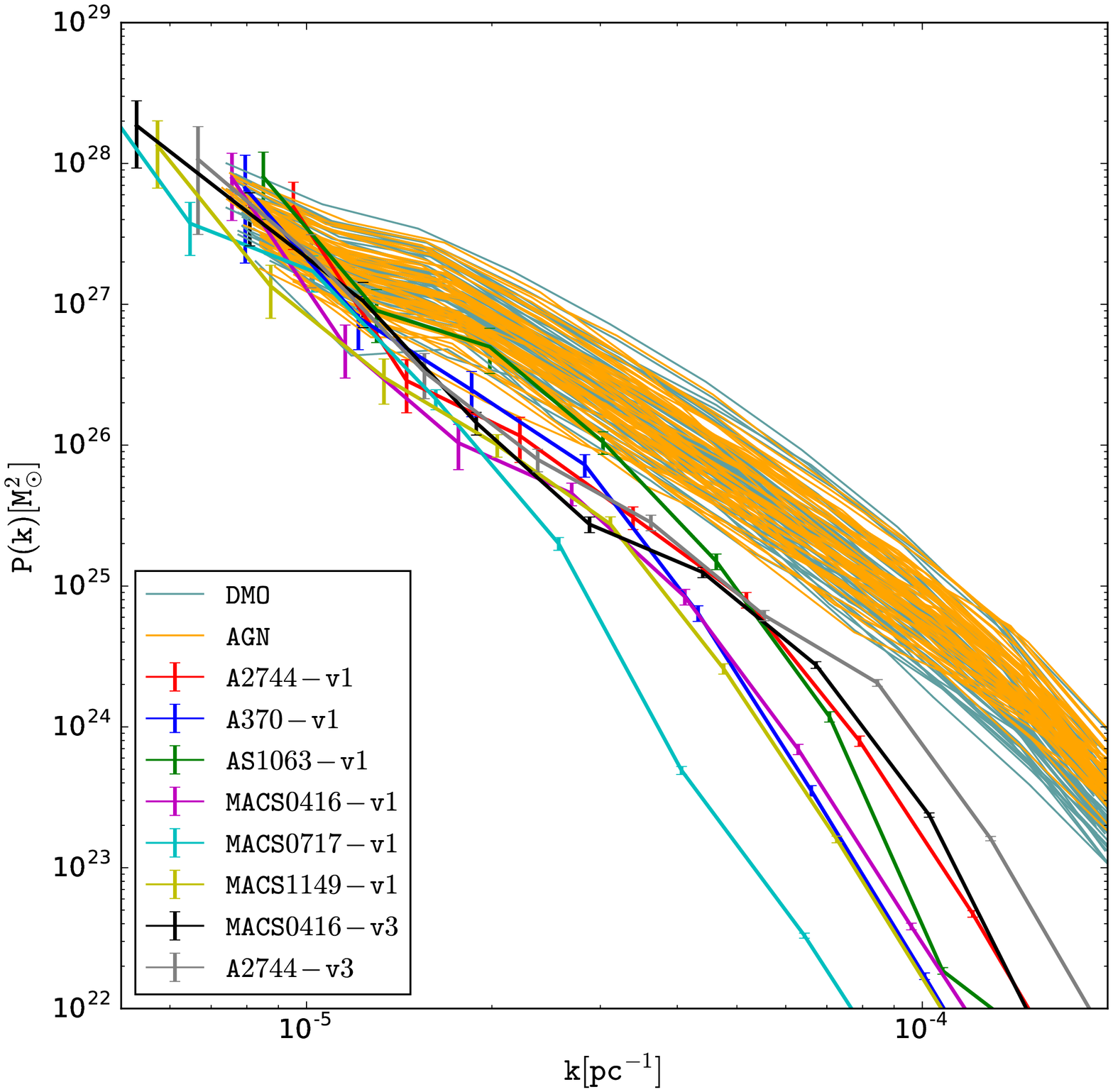}
  \includegraphics[width=0.48\textwidth]{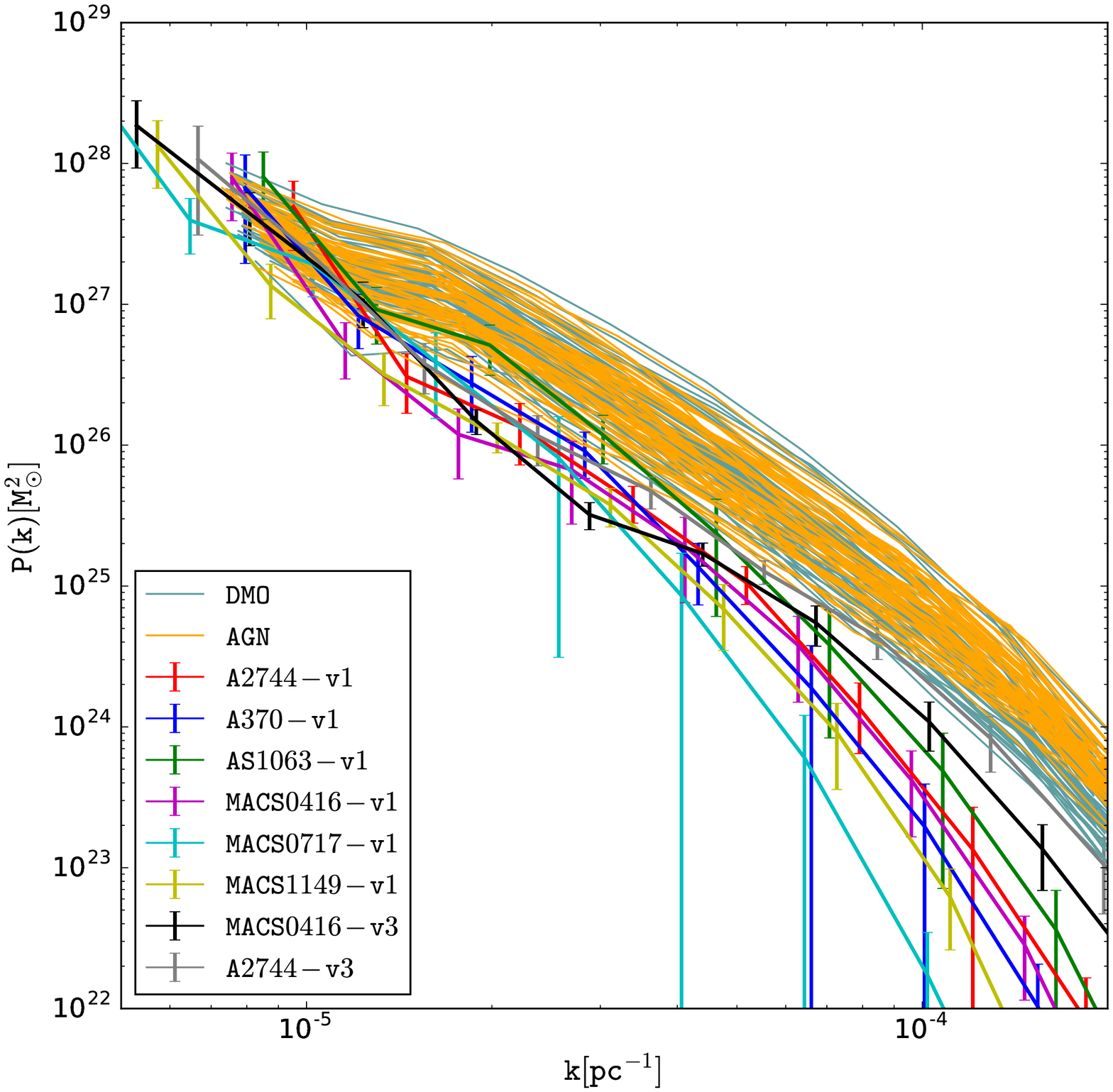}
	\caption{All power spectra: clusters in thick solid lines and
	simulations in thin solid lines.
  Left: showing power spectra calculated for the average mass map of each
  cluster. The error bars are the scatter in different modes at a given $k$
  (or the sample variance).
  Right: each power spectrum is the average of 30 (40 in case of $\mathtt{v3}$) different power
  spectra calculated for each mass map in the ensemble for the respective
  cluster.
  The error bars are the standard deviation of the power spectra,
  including contributions from sample variance, as well as the scatter between different
	reconstructions of the same cluster.}
	\label{fig:pkall}
\end{figure*}

\begin{figure*}
	\centering
	\includegraphics[width=0.48\textwidth]{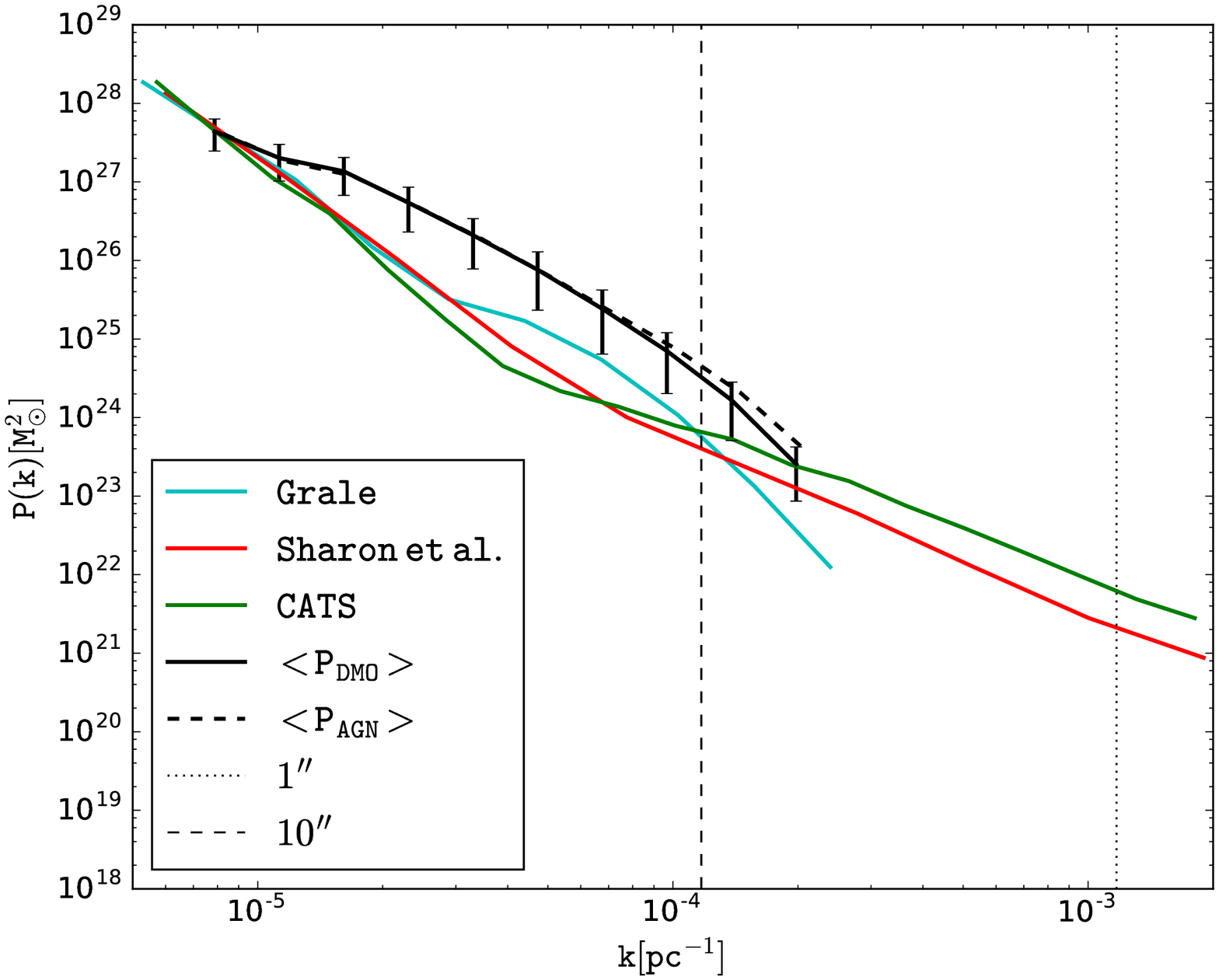}
	\includegraphics[width=0.48\textwidth]{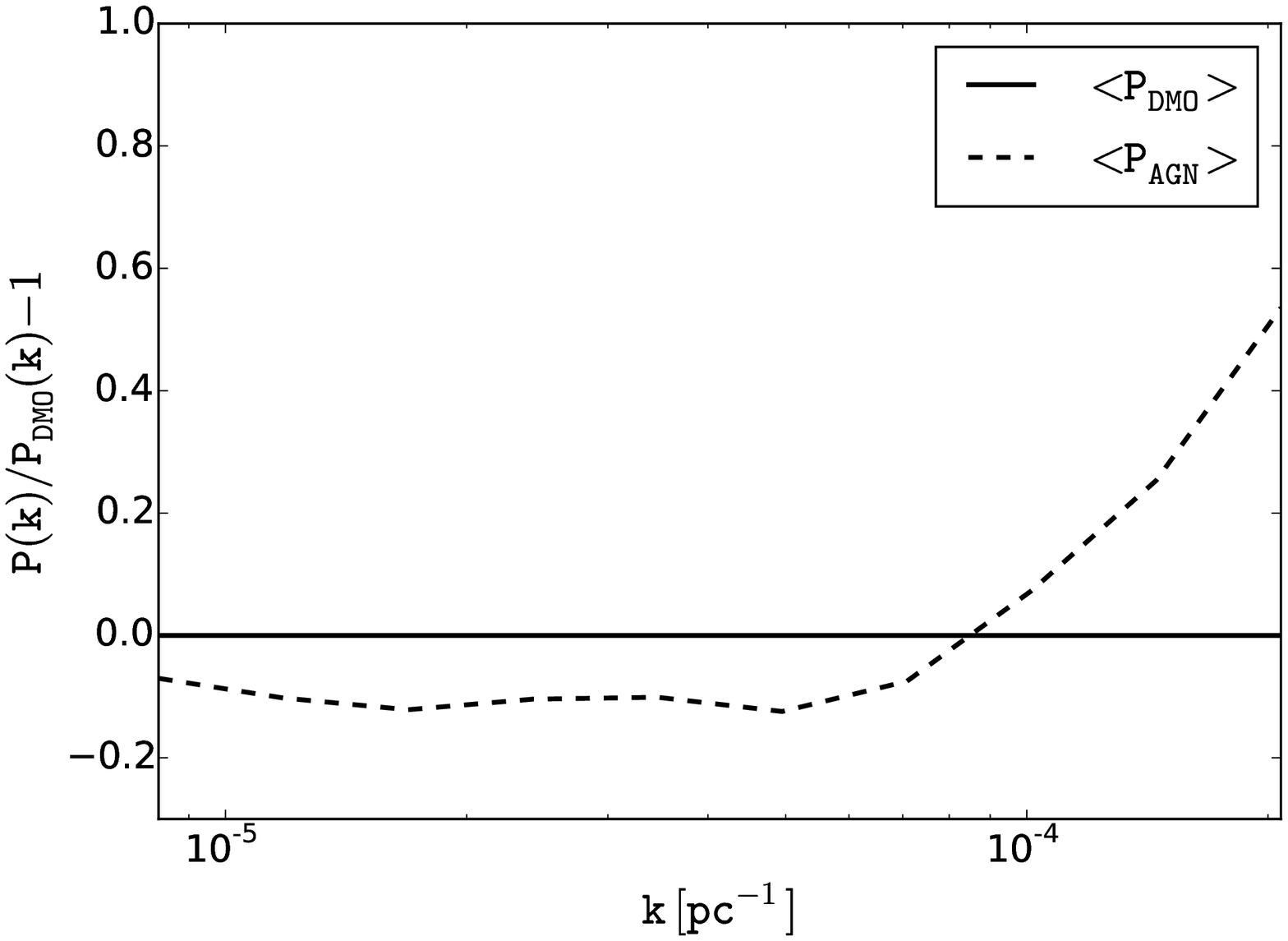}
	\caption{Left: comparing the mean power spectrum for dark-matter only simulations (DMO),
	baryonic simulation (AGN) and different reconstructions using GRALE and LENSTOOL for MACS0416-v3
	(using HFF data only). The vertical lines show the 10 and 1 arc second scales at the cluster's redshift
	$z=0.396$. Right: the relative difference between the
	mean power spectra of DMO and AGN simulations.}
	\label{fig:pk:compare}
\end{figure*}

\begin{figure*}
	\centering
	\includegraphics[width=0.8\textwidth]{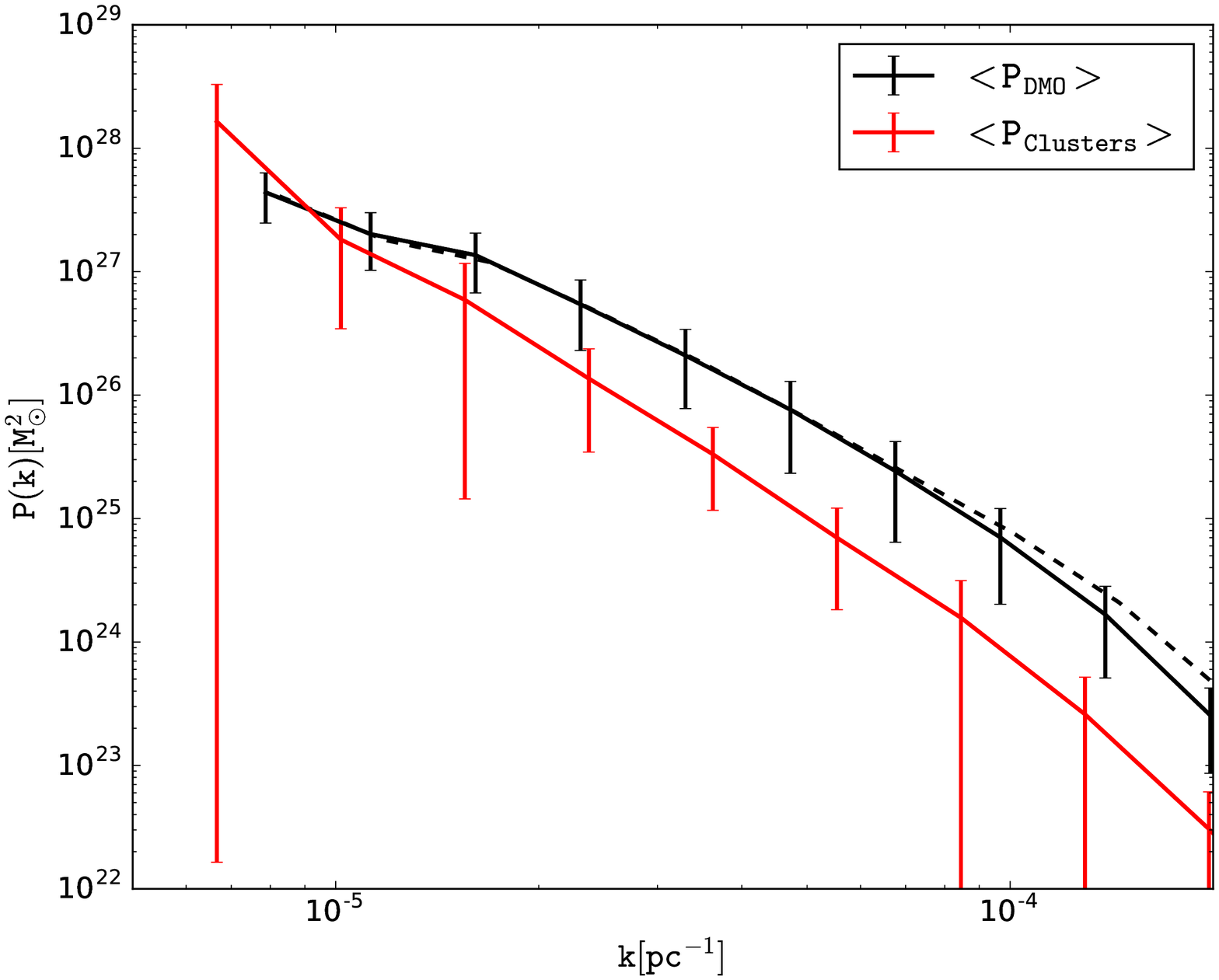}
	\caption{Comparing average power spectrum of HFF clusters to
				that of the simulated clusters.}
	\label{fig:pk:sim}
\end{figure*}

\subsection{Comparing HFF clusters}

We can now compare the clustering properties of HFF clusters
using the lensing power spectrum $P_{\rm M}(k)$ .
The reconstructed mass distributions of the HFF clusters are morphologically different but
show similar statistical structures which is very much evident from comparing the respective
power spectra. Figure \ref{fig:pkall} shows all the power spectra for HFF clusters in
thick lines along with their sample variance, which represents scatter across
all available modes for a given $k$.

Let us first consider two reconstructions for MACS0416, using pre-HFF and HFF data. Both
mass maps are very similar at large scales but differ in detail at small scales. The
same effect can be noticed in the power spectrum (see figure \ref{fig:pkall}): up to
$k \sim 3 \times 10^{-5} \rm{pc}^{-1}$ both look nearly the same, but at smaller scales
the mass map reconstructed with pre-HFF data starts to lose power. This is a
consequence of the fact that HFF data has over two times as many lensed images as pre-HFF
data (see table \ref{tbl:HFF}), allowing for higher resolution in the reconstructed mass maps.
This leads to a higher power at small scales, where the difference in the two power spectra
is almost an order of magnitude. A similar trend can be seen in the two versions
of Abell 2744. 

Power spectra for Abell 370 and MACS1149 look very similar to each other. If we look at the corresponding mass maps, both show two dominant peaks with similar width, shallow as compared to the surroundings with similar smoothing.
Similar arguments can be made for Abell 2744 and MACS0416-II clusters.
The clusters show similar power at small scales which is expected from the very steep mass peaks in the two clusters.

It is important to note that the smallest scale (largest $k$) to be trusted depends on the density
of images. For example, MACS0416-v3 has 4 times more images than MACS0717, (assuming
roughly the same area for both in terms of arcsec$^2$), so the typical linear image
spacing is 2 times larger in MACS0717, leading to poorer mass
resolution. And, in fact,
looking at figure \ref{fig:pkall}, the $k$ value where $P_{\rm M}(k)$  of MACS0717 begins to drop
($k\sim 2\times 10^{-5} {\rm pc}^{-1}$) is about a
factor of $\sim 2$ smaller compared to that of MACS0416
($k\sim 4\times 10^{-5} {\rm pc}^{-1}$).

There is also a tendency for low redshift clusters to have higher power at
small scales. Thus, Abell 2744 reconstructed with 41 images has larger power
than MACS0416-v1 (40 images) at small scales.
Similarly, MACS1149 has less power at small scales than A370 and AS1063, again with similar numbers of lensed images.
This is also intuitive as low redshift clusters have had more time to
build small-scale substructures.
Therefore, the power at small scales may be indicative of the age of the cluster.

\subsection{$\Lambda$CDM simulation clusters}

We used 22 clusters of galaxies from dark-matter only (DMO) simulations and
an additional 22 from hydrodynamical simulations which include AGN feedback,
in the mass range  1--3$\times 10^{14} M_{\odot}$ from \cite{2014MNRAS.440.2290M}
using RAMSES \citep{2002A&A...385..337T}.
The mass range is chosen to be comparable to that of the HFF clusters but the
simulated clusters are all at redshift zero. We then projected
them along the three axes, for a total of 66 projected clusters in each case.
Projection does introduce some correlation in the sample, but we assume
such correlations in the sample, but we neglect this effect.
In contrast to HFF clusters, the simulation clusters are virialised but a few show more than one core.
The measured power spectra are shown as thin lines in figure \ref{fig:pkall}.

We also measured the mean power spectrum in each case, DMO and AGN, which is shown in black
in the left panel of figure \ref{fig:pk:compare}.
The error bars show the standard deviation of $P_{\rm M}(k)$, including contributions from
scatter in different modes at a given $k$ as well as the scatter between different clusters.
In the right panel of figure \ref{fig:pk:compare}, we show
the relative difference between the mean power in DMO and AGN simulated clusters.
Comparing the mean power spectra of DMO and AGN, the latter shows a deficit in power
at large scales and a boost at smallest scales. The deficit is nearly $10\%$.
The AGN clusters have the influence of an AGN at the centre that drives the gas
outside the cluster with its feedback process. Losing this mass results in the
deficit of power at those scales. However, these AGN clusters contain a central
stellar component in the form of the brightest cluster galaxy (BCG),
which increases the mass at the very centre of the
cluster resulting in a boost in the power at smallest scales.
The trend is systematically consistent with previous findings.

\subsection{Comparison between HFF clusters and $\Lambda$CDM simulation clusters}

In figure \ref{fig:pkall}, we show $P_{\rm M}(k)$ for all six HFF clusters as
well as both DMO and AGN for all the simulated clusters.
The lensing clusters systematically show less power at small scales
compared to the simulated ones.
We discuss the possible reasons for this in
section \ref{sec:discussion}.

\subsection{Comparison with LTM models}

In the left panel of figure \ref{fig:pk:compare}, we compare the power spectrum measured
from the mass maps reconstructed using GRALE to that made using other methods:
Sharon et al and the CATS group.\footnote{\tt https://archive.stsci.edu/prepds/frontier/lensmodels/}
Both use the LTM method LENSTOOL of \cite{2007NJPh....9..447J} as the reconstruction technique.
Both of these reconstructions show much larger
power at small scales compared to the GRALE reconstruction. This is the result of the fact that LENSTOOL uses
information from the individual cluster galaxies of the lens, and
hence has much steeper density gradients at small scales which results in
larger power. This effect is also seen in the galaxy-mass correlation function of
MACS0416 which shows a sharp peak at zero separation \citep{2015arXiv150708960S}.
The power spectrum of parametrically reconstructed clusters is a mixture
of priors and lensing information, and when the number of images is small, priors
are the dominant contribution.
In contrast, free-form methods like GRALE base the mass
distribution on the lensed images alone, without relying on visible galaxies.


\section{Conclusions and Discussions}\label{sec:discussion}

In this article, we present free-form mass models of six Hubble Frontier Field
clusters using strong gravitational lensing pre-HFF data.
We also studied two of the six clusters, MACS0416 and Abell 2744, using both pre-HFF
and HFF data (with many more lensed images).
For each cluster, a total of 30 mass maps (40 for $\mathtt{v3}$ versions) are made
using a non-parametric lens inversion technique called GRALE, and the
average mass map is presented. All the mass
maps show elongation, multiple cores and many substructures in the mass
distribution implying a recent major merger. The lensing data from
recent HST-HFF observations is rich in lensing images allowing a very
precise identification of substructures to be done with greater
confidence. These mass maps can be used to study various aspects
of the structure formation and merging stages of large collapsed clusters
of galaxies.

We measured the power spectra of these sky-projected mass maps (30 for each cluster),
and presented both average power spectra, as well as the power spectra of the
average mass map for each cluster (figure \ref{fig:pkall}). The error
estimates of the power spectra are either given as the scatter in different modes
at a given $k$, also known as sample variance, or a combination of the sample
variance and the standard deviation of the power spectra for the same cluster.
Power spectra give a statistical description of the clustering properties of the
mass distribution and encode information about the abundance of
substructures  and their contrast with the background.
There are tentative indications that low redshift clusters
have systematically higher power at small scales as compared
to high redshift clusters, presumably because low redshift
clusters are closer to their virial equilibrium than high redshift clusters.
We argue---and illustrate using pre-HFF and HFF maps of MACS0416---that using a
larger number of lensed images in a cluster leads to a better-constrained mass
distribution, and that the power spectrum
can be recovered up to much smaller scales putting stronger constraints
on our understanding of the substructures. On the other hand, parametric
methods that explicitly include cluster galaxies in the models have much larger power on small scales
even with fewer lensed images indicating that the mass maps are dominated
by the priors, especially at small scales.

We compared the power spectra of the HFF clusters with those from $\Lambda$CDM
dark-matter only and hydrodynamical simulations at redshift zero.
Figure \ref{fig:pk:sim} summarizes our comparison between HFF clusters and
$\Lambda$CDM simulation clusters. We average the power spectra of the six HFF
clusters and compare them with the average power spectra of the simulated
clusters. The average power of the clusters is systematically steeper than
that of the simulated clusters and hence have less power at small
scales. This may be due to one or more of the following reasons:

\begin{itemize}
    \item The redshift of the HFF clusters are higher than the simulated clusters.
    We are comparing non-virialised HFF clusters with the virialised
    halos of the simulations. At later times, the mass distribution
    becomes more and more clumpy and the substructures pull
    more mass from the background,
    which results in a higher power at small scales. Therefore, it
    is possible that the lensing clusters in the local Universe may
    show similar power as the simulated clusters
    at redshift zero. Conversely, this can be checked by comparing
    the power spectrum of HFF clusters with that of simulated clusters of similar masses
    in the redshift range 0.3--0.5.

    Another effect of the redshift difference is that the angular resolution is lower for the higher redshift clusters. However, this effect is roughly 20$\%$, too small to account for the observed differences in the power spectrum.

    \item A second possible reason relates to the data and method we are
    using to reconstruct the clusters. A larger number of lensed images gives
    additional constraints on the mass distribution, and this increased resolution leads to a boost
    in the power spectrum. This can be seen in figure \ref{fig:pkall}  where we show the power
    spectrum of MACS0416 using both pre-HFF and HFF data: HFF data shows
    larger power at small scales as compared to pre-HFF data.

    Figure \ref{fig:pk:compare} shows that the power at small scales
    also depends on the assumption of the reconstruction method.
    It is possible that GRALE does not have enough resolution  and hence loses
    power at smallest scales. On the other hand, LENSTOOL maps seem to have
    much more power than expected at those scales and are possibly dominated by priors.

    \item Finally there is the possibility that the simulations are lacking
    some physics that needs to be taken into account in order to simulate
    realistic halos.
\end{itemize}

The second item above can be tested through a more extended pipeline, in which clusters are generated from
state-of-the-art simulations, lensing data are generated from them, and then analysed by independent groups
using different techniques.  Such tests are currently in progress.
It will be interesting to measure the power
spectrum of the original mass distribution and the reconstructed one
using different lens inversion methods. 
The current analysis expects that the power spectrum based on GRALE maps
will match the true power spectrum at large scales and ultimately
lose power depending on the density of lensed images, whereas
LTM-based methods may continue having larger power even when the true power
drops. We leave this analysis for future work.

Independent of lensing, it will also be interesting to study the evolution of the power spectrum
as a probe of sub-structures and the merging history of large collapsed objects in simulations.
For example, a cosmological simulation can be set up with a volume large enough to produce
20--50 clusters of galaxies,  in a narrow mass range.  The power spectra can be calculated
for all the clusters in the mass range at different redshifts, and the evolution of the average
power spectrum can be studied. This will give us insight into the evolution of clustering
properties in a merging system. We leave this analysis to future work as well.



\section{Acknowledgement}

We would like to thank Davide Martizzi for providing the projected mass maps of the simulated clusters. 
IM would also like to thank Aurel Schneider for
useful discussions about the topic.

LLRW is grateful to the Minnesota Supercomputing Institute for their computational
resources and support.

We are grateful to the numbers of the HFF Map Maker teams, and especially to Dan
Coe, Johan Richard and Anton Koekemoer, who made their image identifications and
redshifts available to us, in some cases prior to publications for the v1 maps. Some
redshifts used were spectroscopic, some photometric and some were predicted by the
lens models. The relevant papers from which image information was taken are cited in
the text.
For v3 models, we are grateful to all those who contributed to the data used here:
members of the CATS team (including Mathilde Jauzac, Johan Richard, Jean-Paul
Kneib), GLAFIC team, Benjamin Clement, Adi Zitrin, Irene Sendra, Claudio Grillo,
Daniel Lam, Jose Diego, Xin Wang, Marusa Bradac, Austin Hoag, Traci Johnson, Brian
Siana, Anahita Alavi, Gabe Brammer, Tomasso Treu, Ryota Kawamata, and especially Dan
Coe and Keren Sharon.

IM is supported by Fermi Research Alliance, LLC under Contract No. De-AC02-07CH11359 with the United States Department of Energy.


\bibliographystyle{mn2e}


\def\apj{ApJ}
\def\apjl{ApJL}
\def\apjs{ApJS}
\def\aj{AJ}
\def\aapr{aapr}
\def\pasj{pasj}
\def\na{na}
\def\araa{araa}
\def\physrep{physrep}
\def\nat{Nature}
\def\mnras{MNRAS}
\def\aap{A\&A}
\def\prd{PRD}

\bibliography{ms}


\end{document}